\documentclass[review,5p]{elsarticle}
\journal{Computational Materials Science}
\usepackage{hyperref}
\usepackage[utf8]{inputenc}
\usepackage{amsmath}
\usepackage{amssymb}

\usepackage{graphicx}
\usepackage[american]{babel}
\usepackage[amssymb]{SIunits}
\usepackage{subfigure}
\usepackage{notoccite}
\usepackage{afterpage}
\usepackage{tikz}
\usepackage{ulem}
\usetikzlibrary{calc}
\usetikzlibrary{plotmarks}
\usepackage{pgfplots}
\usepackage{etoolbox}
\patchcmd{\normalsize}{13.6}{13}{}{}
\pgfplotsset{
tick label style={font=\tiny},
label style={font=\small},
legend style={font=\tiny}
}
\begin{document}

\begin{frontmatter}
 
\title{Eutectic colony formation in systems with interfacial energy anisotropy: A phase field study}

\author[add]{Arka Lahiri\corref{correspondingauthor}}
\ead{arka@platinum.materials.iisc.ernet.in}

\author[add]{Chandrashekhar Tiwary}
\ead{cst311@gmail.com}

\author[add]{Kamanio Chattopadhyay}
\ead{kamanio@materials.iisc.ernet.in}
 
\author[add]{Abhik Choudhury}%
\ead{abhiknc@materials.iisc.ernet.in}

\address[add]{%
Department of Materials Engineering, 
Indian Institute of Science, Bangalore - 560 012, India.
}%
\cortext[correspondingauthor]{Corresponding author}


\begin{abstract}
Instability of a binary eutectic solidification front to morphological perturbations due to rejection of a ternary impurity
leads to the formation of eutectic colonies. Whereas, the instability dynamics and the resultant microstructural features
are reasonably well understood for isotropic systems, several experimental observations point to the existence of 
colonies in systems with anisotropic interfaces.
In this study, we extend the understanding of eutectic colonies to anisotropic systems, where 
certain orientations of the solid-liquid or solid-solid interfaces are associated with a lower free energy than the others. 
Through phase field simulations in 2D and 3D, we have systematically
probed the colony formation dynamics and the resulting microstructures, as functions of the 
pulling velocity and the relative orientation
of the equilibrium interfaces with that of the imposed temperature gradient. 
We find that in 2D, stabler finger spacings are selected with an increase in the 
magnitude of anisotropy introduced, either in the solid-liquid 
or in the solid-solid interface. The fingers have a well-defined orientation 
for the case of anisotropy in the solid-liquid interface, with no fixed orientations for the lamellae 
constituting the colony. For the case where anisotropy exists in the solid-solid 
interface, the lamellae tend to orient themselves along the direction of the imposed temperature gradient, 
with tilted solid-liquid interfaces from the horizontal. The 3D simulations reveal 
existence of eutectic spirals which might become 
tilted under certain orientations of the equilibrium interfaces. Our simulations are able to
explain several key features observed in our experimental studies of solidification in Ni-Al-Zr alloy.  
\end{abstract}

\end{frontmatter}

\section{Introduction}
Two-phase growth in a ternary alloy where the two phases exchange two 
components and reject an impurity into the liquid, results in  
the formation of a boundary layer of this component ahead of the solidification
front. This interface is then unstable to morphological perturbations
much in the same manner as in a Mullins-Sekerka instability~\cite{Mullins1964} (henceforth will be referred to as MS instability)
of the solidification front
during single phase growth. The amplification of these instabilities leads to the formation of eutectic colonies (also called 
two-phase fingers) which are cells made up of two-phase lamellae.

Experimentally, eutectic colonies 
have been extensively studied, as in~\cite{Chilton1961, Kraft1961,Hunt1966,Rumball1968,Rohatgi1969,Bullock1971,Rinaldi1972}. 
including a study on their dynamics during directional solidification of thin samples~\cite{Akamatsu2000}.
Furthermore, Akamatsu et al.~\cite{Akamatsu2010}
are the first to observe and characterize the helical 
arrangement of two eutectic solids about a finger axis, which they anoint as ``eutectic spirals'' . 
They also point out the presence of such structures in studies 
that predate their observation, like the one in Al-Cr-Nb systems carried out by 
Souza et al.~\cite{Souza2005}. 

Theoretical understanding of this problem begin with  
the study by Plapp and Karma~\cite{Plapp1999}, where 
they perform linear stability analysis to establish that the instability 
leading to colony formation is oscillatory compared to the 
one operating on a single phase binary solid-liquid interface. 
% Liu et al. present a theory in~\cite{Liu2009} to 
% examine the modifications of the steady-state 
% growth dynamics of a binary eutectic caused by the addition of a third element.
The experimental observation that a spiral tip radius 
($\rho$) scales linearly with the lamellar width ($\lambda$)~\cite{Akamatsu2010} 
leads to an analytical establishment of the scaling of $\rho$ with 
$V^{-0.5}$, where $V$ is the spiraling dendrite tip growth 
velocity~\cite{Akamatsu2014}.

Numerical computations performed to study eutectic colony formation dynamics 
augment theoretical understanding in regimes outside 
the purview of linear analysis. In this regard, phase field simulations 
of eutectic colonies by Plapp and Karma~\cite{Plapp2002} 
not only validate their theory in~\cite{Plapp1999}, 
but also highlight the lack of a stable cellular morphology 
under isotropic conditions. 
% Recent studies~\cite{Pusztai2013,Ratkai2015} 
% report the stabilisaton of cells in phase field simulations 
% by the introduction of kinetic anisotropy leading to eutectic dendrites 
% resembling 3D spirals as seen in~\cite{Akamatsu2010}. 

While the studies in the previous cases concentrate on 
isotropic eutectics, alloys systems in general contain phases
which either have anisotropic solid-liquid interfaces 
or where the interfacial boundaries between the solid phases have 
a preferred alignment of crystallographic planes
giving rise to defined orientation relationships 
for these interfaces. Experimentally, anisotropic interfacial energies
of the solid-solid interfaces have been found to result in spirals
in binary eutectic alloys~\cite{Fullman1954,Liu1992}. 

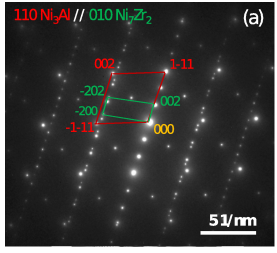
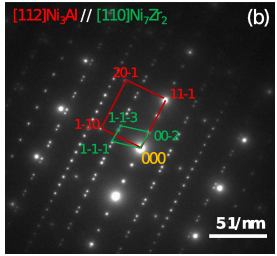
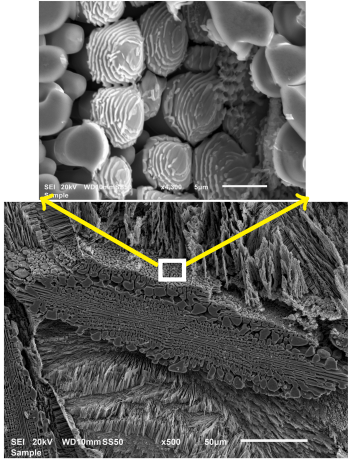
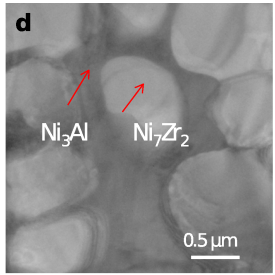
\begin{figure}[!htbp]
\centering
\subfigure[]{\includegraphics[width=0.4\linewidth]{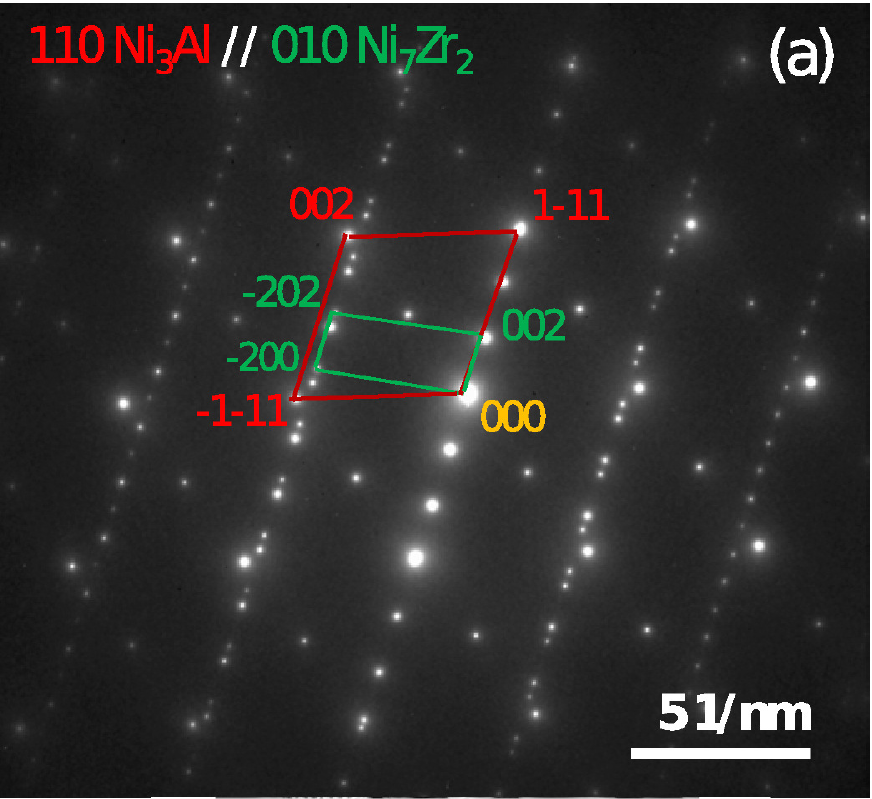}
 \label{tem1_img}
 }
 
  \subfigure[]{\includegraphics[width=0.4\linewidth]{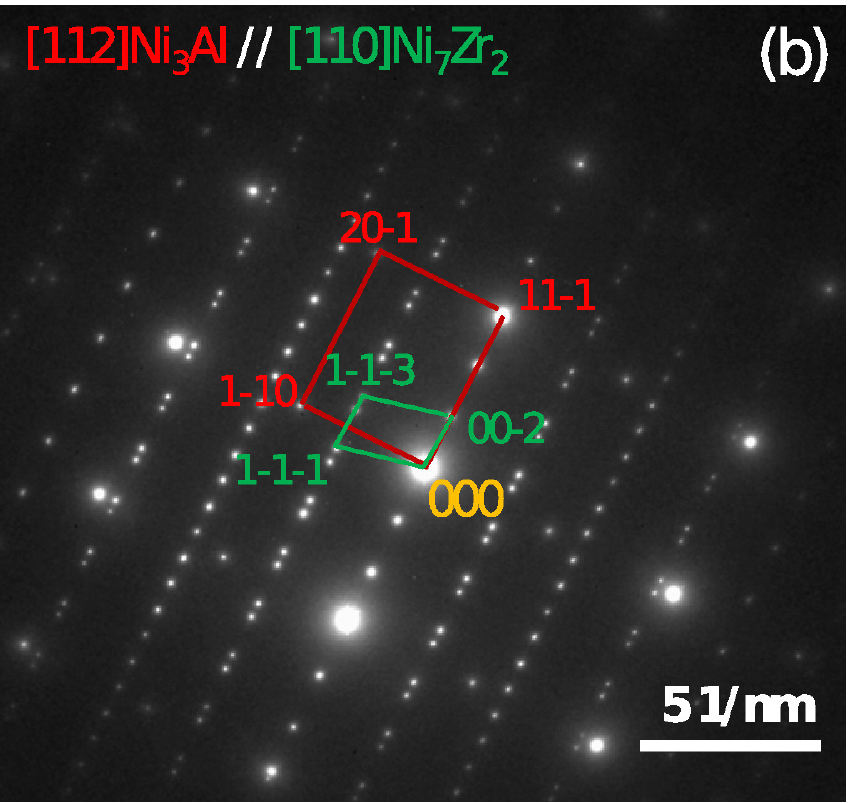}
 \label{tem2_img}
 }
 
 \subfigure[]{
 \begin{tikzpicture}
  \node[anchor=south west,inner sep=0] (image) at (0,0){ \includegraphics[width=0.7\linewidth]{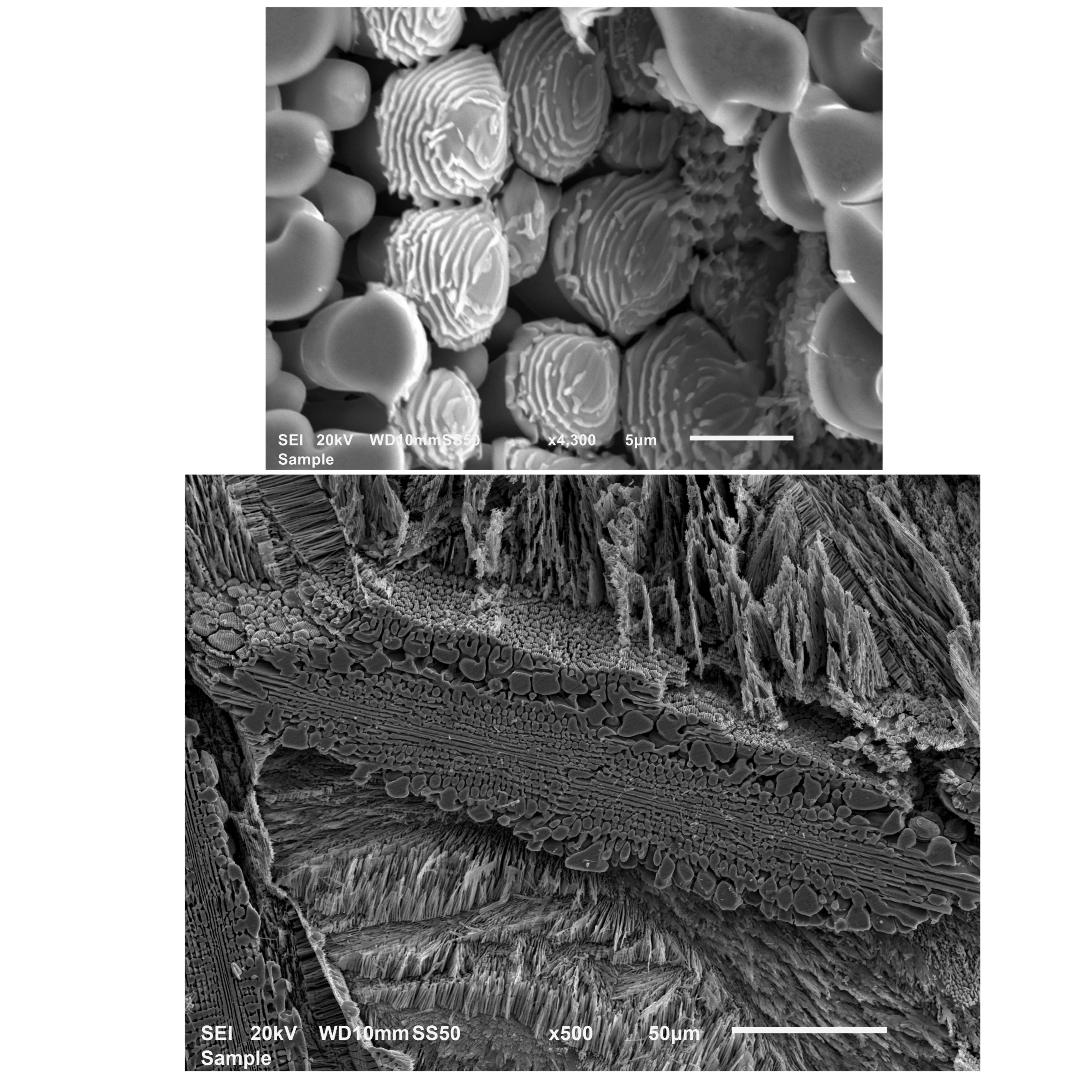}};
  \begin{scope}[x={(image.south east)},y={(image.north west)}]
   \draw [->,ultra thick,yellow] (0.495,0.43) --(0.245,0.575);
    \draw [->,ultra thick,yellow] (0.55,0.43) --(0.815,0.575); 
   \draw [white, ultra thick] (0.5,0.4) rectangle (0.55,0.44); 
   \end{scope}
   \end{tikzpicture}
   \label{finger}
  }
  
  \subfigure[]{\includegraphics[width=0.4\linewidth]{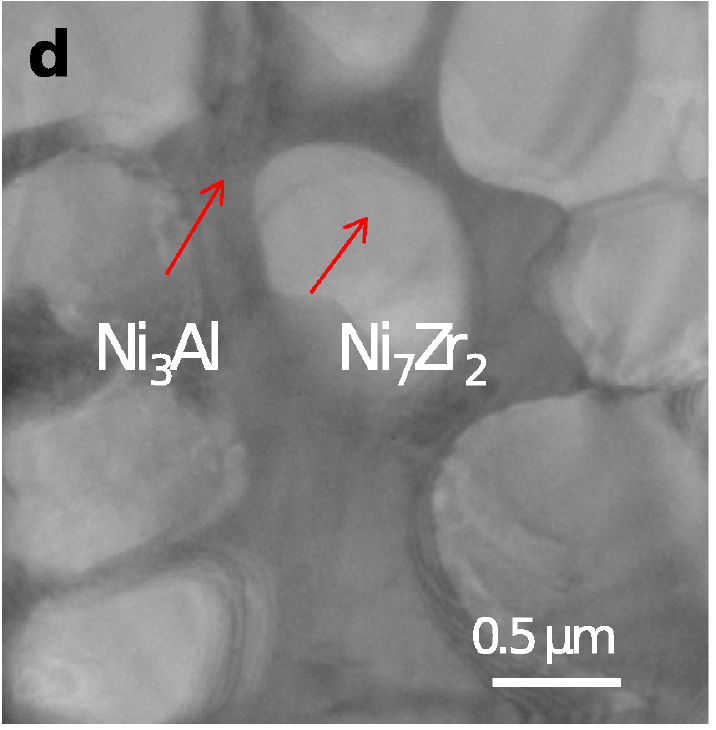}
 \label{fourth_img}
 }
 \caption{The anisotropy in solid–solid interfacial energy is indicated by the existence of a well defined orientation relationship
 between the two phases in (a) and (b). The orientation of the lamellae along the axis of the colony finger along 
 with some spiral like features is displayed in (c). The two eutectic phases are identified from their contrast in (d).}
 \label{exp1}
\end{figure}

Ni-Al-Zr is another exemplary system consisting of two 
solid intermetallic phases ($Ni_3Al$ and $Ni_7 Zr_2$) 
whose crystallographic planes share a well defined orientation 
relationship as revealed by the TEM diffraction patterns in Fig.~\ref{tem1_img} and~\ref{tem2_img}.
The two-phase eutectic in this alloy is also a monovariant reaction and 
is therefore unstable to morphological perturbations. Detailed 
characterization of the colonies shows that the central stem of 
the colonies have well aligned lamellar feature as seen in Fig.~\ref{finger}.
Further resolution of the colony microstructures at the interface
between two colonies reveals features resembling
spiraling of two solid phases.

The influence of a well defined orientation relationship between the eutectic solids in a binary system
on its steady-state growth morphologies has been studied theoretically, numerically and experimentally in
~\cite{Ghosh2015,Akamatsu2012,Akamatsu2012_2}.  
Pusztai et al.~\cite{Pusztai2013} and Ratkai et al.~\cite{Ratkai2015}, investigate 
the influence of kinetic anisotropy in stabilizing the spiral microstructures 
during two-phase eutectic colony formation in a ternary alloy by conducting phase field simulations.
However, no studies exist which 
systematically investigate the influence of anisotropy in the interfacial
energies on the colony morphology arising out of the destabilization of 
steady-state two-phase growth interface, in either thin-film geometry or during 
bulk solidification. 

This motivates the two principal aims in our paper. Firstly, we perform phase field simulations 
to investigate the influence of the anisotropy of the different interphase interfaces on eutectic colony 
morphologies in thin-film geometry. Secondly, through phase-field simulations
in three-dimensions we characterize the influence of anisotropy both 
in the solid-solid and solid-liquid interfaces on the morphologies of the spirals
and the colony structures.

In what follows, we perform phase-field simulations of the following directionally 
solidifying systems:
one where the interfacial energy is isotropic, followed by 
systems with anisotropic solid-liquid and solid-solid interfacial energies in 2D.
Our simulations in this regard can be thought to be representative of the
solidification experiments carried out for thin samples~\cite{Akamatsu2000}. 
The colony formation dynamics and the resultant lamellar morphologies for each 
of these situations are studied as functions of the anisotropy strength 
and the sample pulling velocity.  

We also perform 3D simulations in order to understand the effect of an 
introduction of a third dimension on the lamellar structures 
in directionally solidified systems with 
anisotropic solid-liquid and solid-solid interfacial energies. Our simulations 
are numerical studies of the eutectic spiraling observed experimentally in~\cite{Akamatsu2010}.
The computational cost involved in these simulations restricts us to a single 
choice in both the pulling velocity and the strength
of anisotropy.

We begin with a discussion of 2D isotropic systems.

\section{2D: Isotropic system}
We begin our discussion with the isotropic system  
where we briefly review the mathematical model developed by Plapp
and Karma~\cite{Plapp2002}. The colony formation 
dynamics and the resulting lamellar and cellular morphologies are also discussed here.

\subsection{Phase-field model}
The two independent components chosen to describe the ternary 
system are $u$ and $\tilde{c}$: $u$ participates in the eutectic reaction by being 
redistributed between the two eutectic solids and $\tilde{c}$ is partitioned 
equally between either of the eutectic solids and the liquid phase 
($K$ being the equilibrium partition coefficient) to set up the Mullins-Sekerka (MS) 
instability~\cite{Mullins1964} during directional 
growth. The solid and liquid free energy densities are given by:

\begin{align}\label{fbulk}
f_{sol} &= \frac{1}{8}(u^{2}-1)^{2} +  (\tilde{c} \ln \tilde{c} - \tilde{c} ) 
- (\ln K) \tilde{c} - \frac{\Delta T}{T_{E}}, \\ \nonumber
f_{liq} &= \frac{1}{2}u^{2} +  (\tilde{c} \ln \tilde{c} - \tilde{c}),
\end{align}
% \begin{equation}\label{fliquid}
% f_{liq}=\frac{1}{2}u^{2} +  (\tilde{c} \ln \tilde{c} - \tilde{c}),
% \end{equation}
where $\Delta T/T_E=(T-T_E)/T_E$ is the scaled and non-dimensionalized undercooling in the system 
with $T_E$ and $T$ denoting the non-dimensional eutectic temperature and the temperature field 
in the system, respectively. 
The equilibrium values of $u$ and $\tilde{c}$ in the solid and the liquid phases 
(i.e., $u_s$, $u_l$, $\tilde{c_s}$ and $\tilde{c_l}$) are computed by solving for a
set of equations mentioned in the Appendix. 

For the $u$ field, solving for the equilibrium phase compositions yield $u_s=\pm{1}$ and
$u_l=0$. This allows identification of the eutectic solids with the solid phase corresponding to $u_s=1$  
named $\alpha$ and the one corresponding to $u_s=-1$ called $\beta$.
 
The temperature profile ($T$) in the Bridgman furnace is given by:
\begin{equation}\label{iso_temp}
 T=T_0 + G (z - Vt),
\end{equation}
where $G$ is the imposed thermal gradient along the vertically 
upward direction, $V$ is the pulling velocity, $t$ is the
time and $z$ is the distance measured 
in a frame attached to the solidification front at $t=0$.
The constant $T_0$ is calculated by setting the 
undercooling at the solid-liquid interface at $t=0$ to a pre-determined value.

The free energy functional representing a solidifying system containing diffuse interfaces is given by~\cite{Cahn1958, Karma1994},
\begin{align}
 F&=\int_V \Bigg[ \left(h(\phi) f_{sol} + (1 - h(\phi)) f_{liq} \right) + \nonumber \\
 &\dfrac{W_u^2}{2} \left(\nabla u\right)^2 
 + \dfrac{W_\phi^2}{2} \left(\nabla \phi\right)^2\Bigg] dV,
 \label{free_func}
\end{align}
where $V$ is the volume undergoing eutectic solidification. $W_u$ and $W_\phi$
are the parameters determining the energy penalty associated with the presence of gradients in $u$ and $\phi$ respectively.
A liquid-to-solid phase transformation is modeled by solving the 
Allen-Cahn equation~\cite{Allen1979} which represents a minimization of $F$ w.r.t 
the spatial variation of $\phi$, where $\phi=0$ denotes liquid and 
$\phi=1$ represents the solid with values between $0$ and $1$ existing 
at the diffuse solid-liquid interface. The governing equation for $\phi$ evolution can be written as,

\begin{equation}
 \tau\frac{\partial \phi}{\partial t}=W_{\phi}^{2} \nabla^{2}\phi - g'(\phi) + h'(\phi)(f_{liq}-f_{sol}),
 \label{fi_evol}
\end{equation}

where $'$ indicate derivatives with respect to $\phi$. 
$\tau$ is the relaxation time for $\phi$ evolution.
The potential barrier between the solid and the liquid phases 
is given by: $g(\phi)=\phi^{2} (1-\phi)^{2}$, and the last term in 
the RHS of Eq.~\ref{fi_evol} represents the driving force for solidification obtained 
from the relative difference in the bulk free energy densities of the solid and the 
liquid phases with the total bulk energy density of the system at any point
in space being given by: 
 \begin{equation}\label{ftotal}
 f = h(\phi) f_{sol} + (1 - h(\phi)) f_{liq},
 \end{equation}
where $h(\phi)=\phi^{2}(3-2\phi)$ is a polynomial interpolant ($h(\phi)=1$ for solid
and $h(\phi)=0$ for liquid).
The evolution of $u$ and $\tilde{c}$ with time are 
obtained by solving the Cahn-Hilliard equation~\cite{Cahn1961} as 
given by:

\begin{equation}\label{u_evol}
 \frac{\partial u}{\partial t} = \nabla \cdot \left[ M \nabla \left(\frac{\partial f}{\partial u} 
 -W_{u}^{2}\nabla^{2} u\right)\right],
\end{equation}
and,
\begin{equation}\label{c_evol}
 \frac{\partial \widetilde{c}}{\partial t}=\nabla \cdot \left[ \widetilde{M} \nabla \left(\frac{\partial f}{\partial \widetilde{c}}
  \right)\right].
\end{equation}
$M$ and $\widetilde{M}$ are the mobilities corresponding to evolution of $u$ and 
$\widetilde{c}$ respectively, which are given by,

\begin{align} \label{mob_form}
 M &= D\left(1-\phi^n\right), \\ \nonumber
 \widetilde{M} &= \widetilde{D}\left(1-\phi^n\right)\widetilde{c},
\end{align}

where $D$ and $\widetilde{D}$ represent constants set to unity. 
Eq.~\ref{mob_form} ensures that there is no diffusion of solutes
inside the solid compared to that in the liquid. As the exchange of $u$ 
between $\alpha$ and $\beta$ happens only
at the advancing solidification front, higher values of the constant $n$ will be required to 
allow for complete solute re-distribution in simulations of directional 
solidification at higher pulling velocities ($V$). 

All of the $\phi$, $u$ and $\widetilde{c}$ 
fields undergo changes across a solid-liquid interface:
$\phi$ changes from $1$ to $0$, $u$ changes from $\pm{1}$ to $0$ and 
$\widetilde{c}$ from $\widetilde{c_s}$ to $\widetilde{c_l}$. So, a solid-liquid 
interface can be isolated from the $\phi$ field by identifying 
locations where it has values lying between $1$ and $0$.   
But across a solid-solid ($\alpha$-$\beta$) interface only 
the $u$ field can be seen to be varying,
(i.e., it changes from $1$ to $-1$ as we go from $\alpha$ to $\beta$) 
while $\phi$ and $\tilde{c}$ remain constant; this provides the only means 
of identifying the solid-solid interfaces.

\subsection{Results}
\begin{figure}[!htbp]
 \centering
 \subfigure[]{\includegraphics[width=0.4\linewidth]{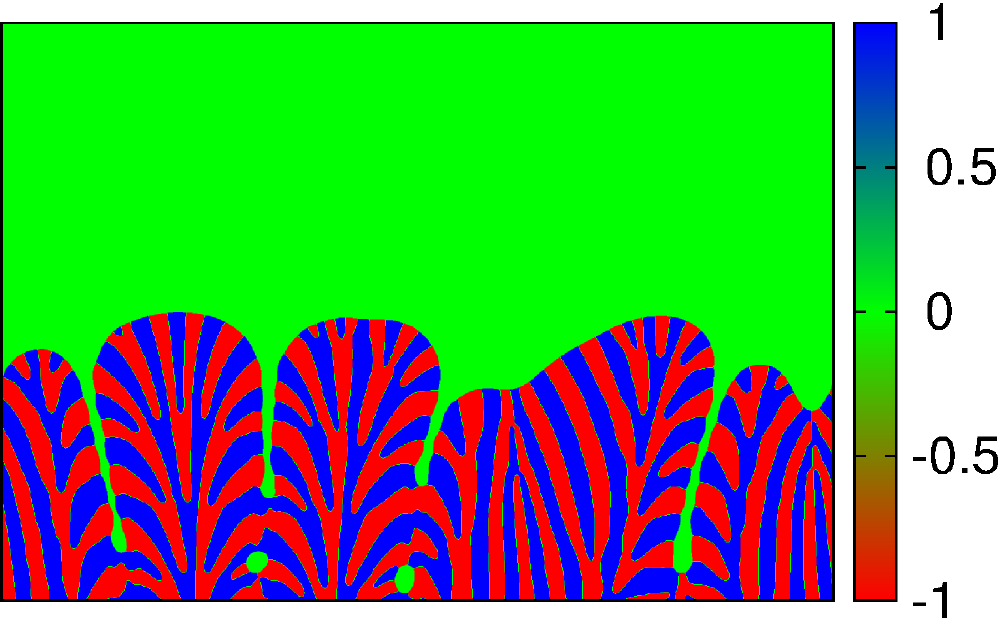}
 \label{iso_2D_u_no_ds}
 }
 %\hspace{.25in}
 \subfigure[]{\includegraphics[width=0.4\linewidth]{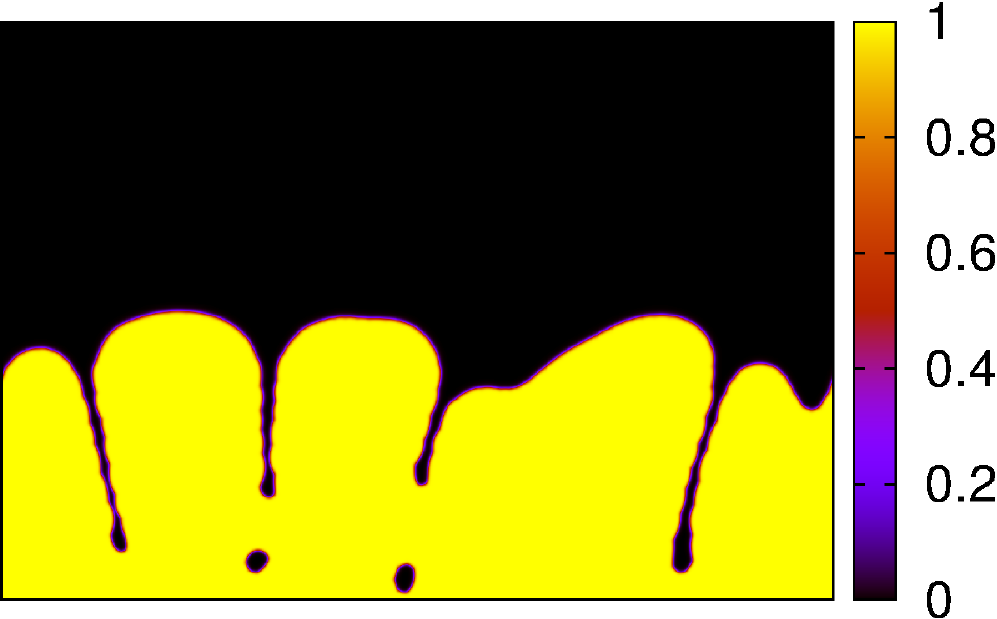}
 \label{iso_2D_fi_no_ds}
 }
\subfigure[]{\includegraphics[width=0.4\linewidth]{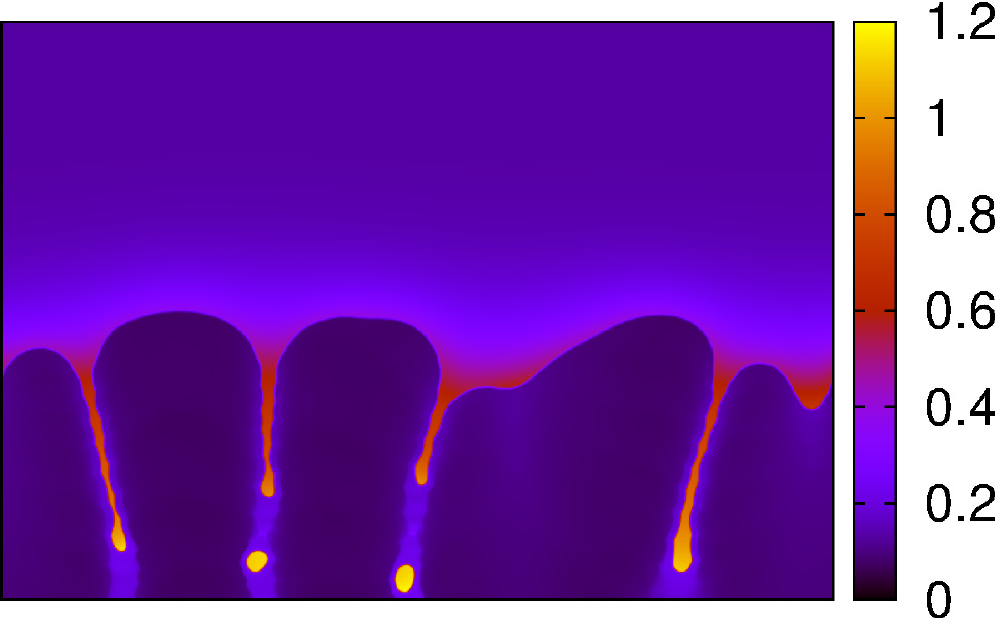}
\label{iso_2D_c_no_ds}
 }
 %\centering
\caption{Colonies in an isotropic system at a total time of 150000 as seen 
from the (a)$u$ field, (b)$\phi$ field, and (c)$\tilde{c}$ field, 
with no diffusivity in the solid. Colorbars report values of the $u$, $\phi$ and $\tilde{c}$ fields 
in (a), (b) and (c) respectively. 
The simulation parameters used for this study are : $G=0.001$, $V=0.015$,
$\tau=1.0$, $D=1$, $\tilde{D}=1$, $n=4$,
$dt=0.0025$, $dx=1$, $dy=1$, $W_{\phi}=3.2$, $W_{u}=1.7$, $\tilde{c_s}=0.025$ and $\tilde{c_l}=0.125$. 
The parameter $T_0$ in Eq.~\ref{iso_temp} is computed by setting the initial undercooling at 
the solid-liquid interface to $0.1$. Periodic boundary conditions are
imposed at the vertical boundaries and no-flux at the horizontal ones in a simulation 
box of dimensions $1440$ by $1000$ containing $40$ lamellae pairs. }
\label{iso_2D_no_ds}
\end{figure}

The 2D simulation of such a system (see Fig.~\ref{iso_2D_no_ds}) 
illustrates the fundamental eutectic cell formation dynamics.
The introduction of some random noise at the solid-liquid interface 
at $t=0$, sets up the MS-type instability through the $\tilde{c}$ component (see 
Fig.~\ref{iso_2D_c_no_ds}). 

The morphological instability of the solid-liquid 
interface leads to formation of fingers, which are identified by 
digit like protuberances of the solid into the liquid 
(see Fig.~\ref{iso_2D_no_ds}). 
Here, the system does not select a particular eutectic
finger spacing and the morphological evolution of the 
fingers display a cyclical character. The fingers that are already formed continue 
to broaden and ultimately develop solid-liquid interface concavities 
which continue to deepen and lead to their splitting, forming new fingers. 
These fingers are randomly oriented with respect to the pulling direction.  

Throughout the course of the simulation, lamellae can be seen to undergo 
either termination or broadening followed by the formation of new lamellae 
by spinodal decomposition. Lamellae pairs converge and ultimately terminate 
at locations where the solid-liquid interface is concave inwards. 
At locations where the solid-liquid interface is convex 
outwards (usually at the tip of the fingers), 
the eutectic phase (either $\alpha$ or $\beta$) present there 
broadens till it develops a concavity at its interface with the liquid, 
prompting formation of the conjugate eutectic solid phase there. It must be
noted at this point, that the mechanism of lamellae broadening followed by 
formation of a new phase by spinodal decomposition, 
prevents finger splitting, which would happen with 
the concavity deepening without a new phase appearing ahead. 

The new lamellae which come out as a result of spinodal decomposition possess a 
lamellar width ($\lambda$) which is different to the one selected by the criterion of 
minimum undercooling at the solid-liquid interface due to Jackson and Hunt~\cite{Jackson1966}. 
Phase separation (the liquid composition $u=0$ decomposing to give $u=\pm 1$ corresponding 
to the two eutectic solids)
happens only at the solidification front ($\phi$ having values between $0$ and $1$) where
the spinodal length scale can be determined by following the analysis in~\cite{Cahn1961},
considering only the double welled part of the potential ($f_{sol}$) in Eq.~\ref{u_evol} as,
\begin{align}
 \frac{\partial u}{\partial t} &=  \nabla \cdot \left[ M \nabla \left(\frac{\partial f_{sol}}{\partial u} 
 -W_{u}^{2}\nabla^{2} u\right)\right] \nonumber \\
 &= \nabla \cdot \left[M\left(\dfrac{\partial^2 f_{sol}}{\partial u^2} \nabla u-W_u^2\nabla\left(\nabla^2 u \right)\right)\right] \nonumber \\
 &= M \left[\dfrac{\partial^2 f_{sol}}{\partial u^2} \nabla^2 u - W_u^2\left(\nabla^4 u \right) \right].
\label{u_evol_mod}
 \end{align}
where we have assumed $M$ to be a constant (but less than $D$) at the interface and retained only linear terms in order to obtain the last equality.
An expression describing the amplification of a sinusoidal variation in $u$($=u_0+A(t)\cos \omega x$), with time,
whose evolution is governed by Eq.~\ref{u_evol_mod}, is given by:
\begin{align}
 A(t) &= A_0\exp\left[R(\omega) t \right] \nonumber \\
 &= A_0 \exp \left[-M \omega^2\left(\dfrac{\partial^2 f_{sol}}{\partial u^2} +  W_u^2 \omega^2 \right) t \right],
 \label{ampl_u_wave}
\end{align}
where $A$ and $A_0$ are the amplitudes at times $t$ and $t=0$ respectively and $\omega$ is the wavenumber of the sinusoidal variation in $u$.
The amplification factor, \\ $R(\omega)=-M \omega^2\left(\partial^2 f_{sol}/\partial u^2 +  W_u^2 \omega^2 \right)$,
has a maximum for a wavenumber of,
\begin{align}
 \omega_{max}= \sqrt{-\dfrac{\dfrac{\partial^2 f_{sol}}{\partial u^2}}{2 W_u^2}},
 \label{omega_max}
\end{align}
which leads us to an expression for $\lambda_{max}^{spin}$(=$2\pi/\omega_{max}$), the dominant length scale 
of spinodal decomposition. Using, the simulation parameters mentioned in the caption to Fig.~\ref{iso_2D_no_ds},
and evaluating $\partial^2 f/\partial u^2$ at $u=0$ we retrieve $\lambda_{max}^{spin}=21.3$. It must be noted at
this point that the dominating wavelength of spinodal decomposition ($\lambda_{max}^{spin}$) is decided only by an 
interplay of bulk and gradient energies as can be seen from Eq.~\ref{omega_max} and is independent of the sample pulling velocity ($V$). 

The other important length scale in the problem is the lamellar width ($\lambda_{JH}$)
corresponding to the minimum undercooling at the eutectic front,
which can be obtained by evaluating the expressions in~\cite{Plapp2002}.
For the parameters of our study, the Gibbs-Thomson coefficient 
evaluates to $\Gamma=1.38$, and the contact angles are $\theta=23.87\degree$, leading to $\lambda_{JH}=40.6$ for $V=0.01$.
Invoking the theory of marginal stability of eutectics~\cite{Langer1980}, all the eutectic length scales
that are smaller than $\lambda_{JH}$, disappear with time under a long wavelength perturbation of the interface,  
leading to an average lamellar width in the system which is larger than $\lambda_{JH}$. This conclusively 
establishes that the microstructural length scales in our simulations are not determined by spinodal decomposition unless at pulling velocities
of $V>0.04$ (from the scaling of $\lambda_{JH}$ with $V^{-0.5}$ given by~\cite{Jackson1966}).
At such high velocities, $\lambda_{JH}$ becomes smaller than $\lambda_{max}^{spin}$ (which remains invariant with change in $V$) 
and the lamellar width set by spinodal decomposition becomes the dominating microstructural length scale in the absence of perturbations
of the solid-liquid interface.   

Thus in 2D, nucleation in this manner
offers only a mechanism to obtain the conjugate phase at a solid-liquid 
concavity. This is however unrealistic given that the undercoolings are not 
sufficiently high for such nucleation to occur. In reality it is a 3D mechanism, 
where a single phase rod rotates to 
appear from planes in front or behind to occupy the concavity. Therefore, 
here we will treat this formation of the second phase as only a mechanism 
allowing one to maintain the scale of the simulation.

Furthermore, in Fig.~\ref{iso_2D_u_no_ds}, there is no specific orientation relationship between the lamellae and the direction of 
solidification (vertically upwards) that is selected by the system.  The lamellae appear oriented roughly orthogonal to the 
solidification envelope (Cahn's hypothesis) close to the solid-liquid interface 
(set by the force balance at the triple points) and take up more random orientations inside the fingers.

By reviewing the isotropic system, we have gained an understanding of the phase-field model which
we are going to build upon in order to study the implications of incorporating anisotropic solid-liquid and solid-solid 
interfacial energies on colony formation. The colony dynamics and morphology observed in the 
2D simulations of isotropic systems also provide a 
reference against which we can attempt to understand the effect of anisotropic interfaces on the lamellar and cellular morphologies,
beginning in the next section.

\section{2D:Effect of anisotropic interfacial energies on the colony dynamics}
In this section, we describe phase-field models of systems possessing anisotropic interfacial energies
and attempt to understand eutectic colony dynamics from 2D simulations. We begin our discussion by considering 
a system with anisotropic solid-liquid interfaces and follow it up with a discussion on systems with anisotropic
solid-solid interfaces. 

\subsection{Anisotropic solid-liquid interface}
In this section we study the cellular features and lamellar orientations in the presence of anisotropic solid-liquid 
interfaces. We draw upon our observations for an isotropic system as a context to understand the simulation results
for this situation. We begin with a description of the phase-field model.

\subsubsection{Phase Field Model}
\begin{figure}[!htbp]
 \centering
  \subfigure[]{\includegraphics[width=0.6\linewidth]{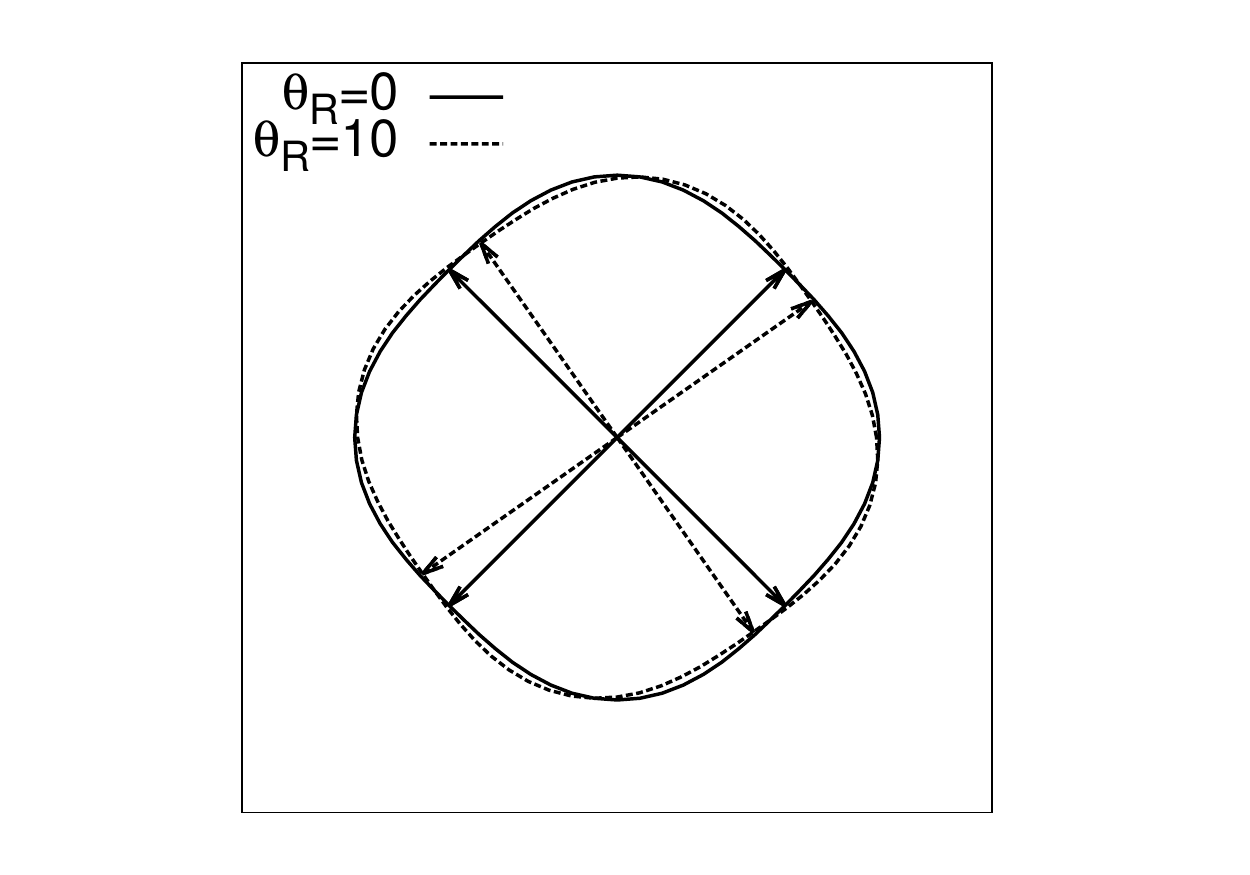}
  \label{gamma_plot}
  }
  \subfigure[]{
  \begin{tikzpicture}
  \node[anchor=south west,inner sep=0] (image) at (0,0){\includegraphics[width=0.4\linewidth]{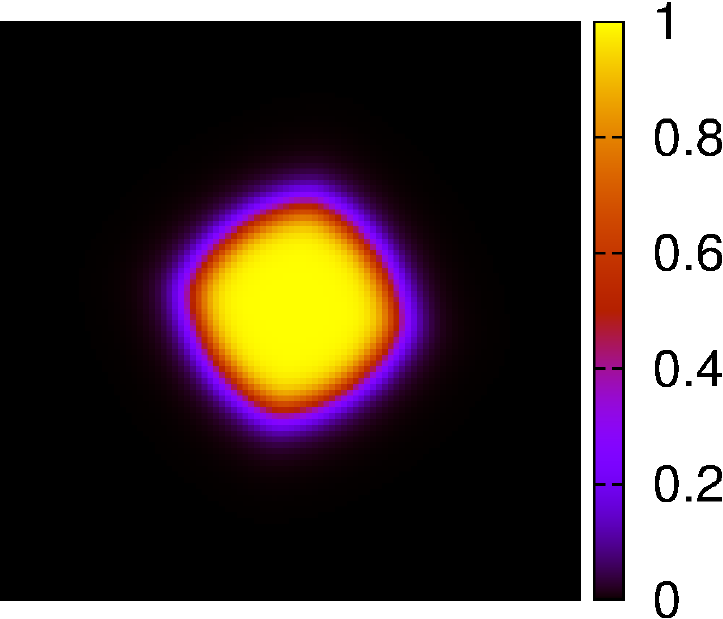}};
  \begin{scope}[x={(image.south east)},y={(image.north west)}]
   \draw [->,ultra thick,white] (0.0,0.53) --(0.25,0.53);
   \draw [->,ultra thick,red] (0.43,0.9) --(0.43,0.7);
   \end{scope}
   \end{tikzpicture}
  \label{rot_10}
  }
  \caption{$\gamma$ plot obtained by evaluating $a_c$ from Eq.~\ref{ac_aniso_fi} is shown in (a).
  The arrows indicate the orientations of the plane normals with the least energy. 
  Figure legends report $\theta_R$ in degrees. (b) A phase field simulation of an 
  $\alpha$ nuclei growing in the liquid, with $\theta_R=10\degree$, clockwise, with $\delta=0.05$.
  The arrows denote the corners which can advance under directional solidification conditions. The corner
  identified by the red arrow dominates over the one indicated by the white one due to its closer 
  alignment to the vertically imposed temperature gradient. Colorbar reports values from the $\phi$ field.}
 \label{orient}
\end{figure}

 A convenient way to identify the solid-liquid interface is with gradients in $\phi$. 
 So, in order to understand the effect of a solid-liquid 
 interfacial energy anisotropy on the microstructural features obtained during directional solidification, 
 we introduce the anisotropy through the gradient energy term in the evolution equation of the $\phi$ field as given by the 
 modified Allen-Cahn equation which writes:
 \begin{equation}
 \tau\frac{\partial \phi}{\partial t}=\left(\nabla \cdot
 \frac{\partial}{\partial \nabla \phi}\right) a - g'(\phi) + h'(\phi)(f_{liq}-f_{sol}),
 \label{aniso_fi_evol}
\end{equation}
where,
\begin{equation}
 a=\frac{1}{2} W_{\phi}^{2} a_{c}^{2}(\theta) (\nabla \phi)^{2}.
 \label{a_aniso_fi}
\end{equation}
The anisotropy function ($a_c$) is given by:
\begin{equation}
 a_c=1 + \delta cos (4(\theta-\theta_R))=
 1-\delta \left(3-4\left(\frac{{\phi^*}_{x}^{4}+{\phi^*}_{y}^{4}}{({\phi^*}_{x}^{2}+{\phi^*}_{y}^{2})^{2}}\right)\right),
\label{ac_aniso_fi}
\end{equation}
which introduces the four-fold anisotropy into the solid-liquid interfacial energy. 
The $^*$'s in the above equation indicate that the derivatives 
(with respect to either $x$ or $y$ as denoted by the subscripts)  
are computed in the reference frame of the crystal. The crystal reference frame can be 
rotated by an angle $\theta_R$ to the laboratory frame and this 
allows us to explore different relative orientations of the equilibrium solid-liquid 
interfaces with respect to the sample pulling direction 
(which is vertically downwards). $\delta$ sets the strength of the anisotropy.
Fig.~\ref{gamma_plot} displays $a_c$ as a function of $\theta$ ($\gamma$ plot), also 
highlighting the effect of a rotation of a crystal frame to the laboratory frame. 

The other equations employed to model a system with anisotropic solid-liquid interfaces 
remain the same as reported in the isotropic situation.

\subsubsection{Results}
The selection of particular orientations of the solid-liquid interface under 
different rotations of crystal frame can also be understood by referring to 
Fig.~\ref{rot_10}.
% \begin{figure}[!htpb]
% \centering
% \subfigure[]{\includegraphics[width=0.4\linewidth]{single_fi_aniso_500_fi_rot_45}
%  \label{rot_45}
%  }
% \subfigure[]{\includegraphics[width=0.4\linewidth]{single_fi_aniso_500_fi_rot_60}
%  \label{rot_60}
%  }
% \caption{$\alpha$ nuclei growing in the liquid, with (a)$\theta_R=45\degree$, and (b)$\theta_R=60\degree$, clockwise, with $\delta=0.05$.
% Colorbar reports values from the $\phi$ field.}
% \label{aniso_fi_single}
% \end{figure}

\begin{figure}[!htbp]
 \centering
  \subfigure[]{\includegraphics[width=0.4\linewidth]{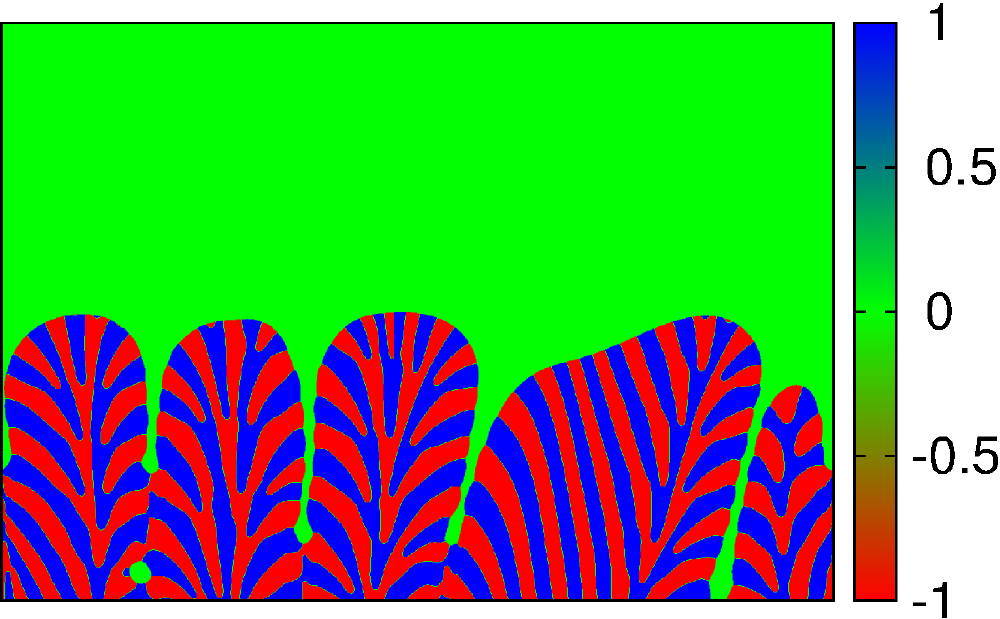}
 \label{ds_0_fi_v_0.015_del_0.015}
 }
 %\hspace{.25in}
  \subfigure[]{
  \begin{tikzpicture}
  \node[anchor=south west,inner sep=0] (image) at (0,0){\includegraphics[width=0.4\linewidth]{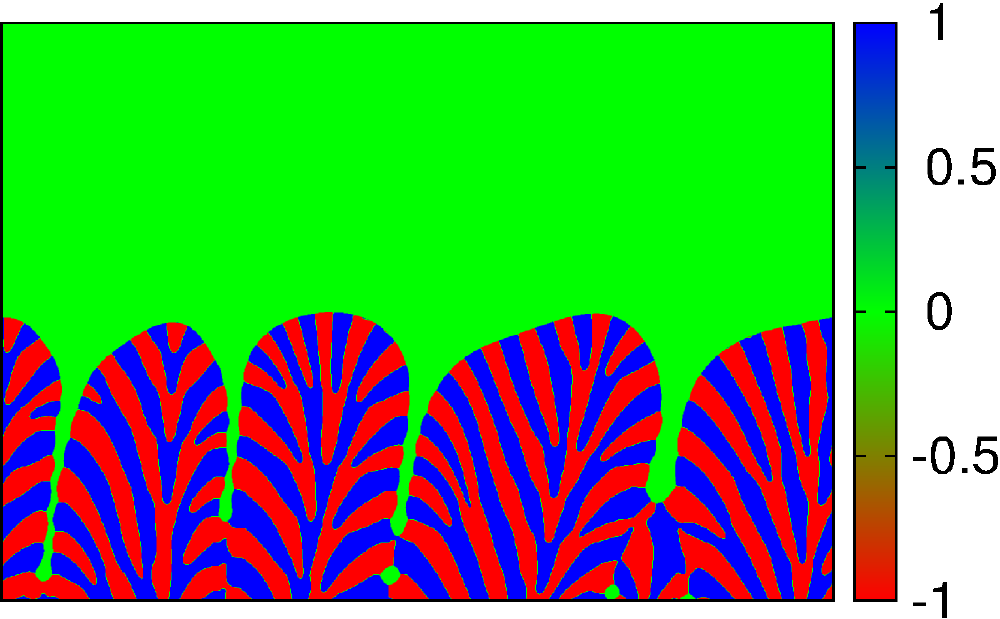}};
  \begin{scope}[x={(image.south east)},y={(image.north west)}]
   \draw [->,ultra thick,yellow] (0.52,0.01) --(0.59,0.51);
   \end{scope}
   \end{tikzpicture}
  \label{ds_0_fi_v_0.015_del_0.03}
 }
\caption{Microstructures ($u$ field) of a system with no solid diffusivity 
and solid-liquid interfacial energy anisotropy, at a total time of $t=150000$, for (a)$\delta=0.015$, 
and (b)$\delta=0.03$, with $V=0.015$, $n=4$ and $\theta_R=10\degree$, clockwise. Colorbars report values of the $u$ field. The other 
simulation parameters are the same as mentioned in the caption to Fig.~\ref{iso_2D_no_ds}.
The arrow roughly indicates the orientation of the finger envelope.}
\label{ds_0_fi_aniso}
\end{figure}

The dynamics of colony formation in such a system is explored for a 
situation where the crystal frame is rotated clockwise
by $\theta_R=10\degree$ to the laboratory frame (see Fig.~\ref{gamma_plot}) for two different 
strengths of the $\phi$ anisotropy, i.e., $\delta=0.015$, and $0.03$, but for 
a single sample pulling velocity ($V=0.015$) (see Fig.~\ref{ds_0_fi_aniso}). 

A lot of features in 
these simulations are in contrast to the isotropic case. First of all,
the finger envelopes tend to favor certain orientations
dictated by an interplay between the anisotropy and the direction of the 
imposed temperature gradient (vertically upward). To illustrate this 
further, we can imagine that the nucleus in
Fig.~\ref{rot_10} being subjected to a temperature gradient prompting it to 
grow in the vertically upward direction. Now, the two corners (indicated by arrows)
in the top half of the crystal are the ones which could grow such 
that the nucleus continues to be bounded by interfaces 
which are favored by anisotropy. But the one on the right 
(identified by the red arrow in Fig.~\ref{rot_10}) 
is usually favored because of it being closer to the pulling direction. 
This can be clearly confirmed in Fig.~\ref{ds_0_fi_v_0.015_del_0.03}
where the fingers have an orientation given by a slight clockwise rotation
from the vertical (represented by the arrow in Fig.~\ref{ds_0_fi_v_0.015_del_0.03}). In situations
where the solid-liquid interface is not as anisotropic 
(lower $\delta$, as in Fig.~\ref{ds_0_fi_v_0.015_del_0.015}), the selection of growth direction is 
not as strict as in the case with higher $\delta$ (see Fig.~\ref{ds_0_fi_v_0.015_del_0.03}).
This manifests as the growth of fingers along directions which 
are not the closest to that suggested by anisotropy under a temperature gradient. Another consequence of this is the 
broader appearance of fingers for $\delta=0.015$ (Fig.~\ref{ds_0_fi_v_0.015_del_0.015}) than for $\delta=0.03$ (Fig.~\ref{ds_0_fi_v_0.015_del_0.03}).

Also, the tilted orientation of the fingers from the vertical as observed in Fig.~\ref{ds_0_fi_v_0.015_del_0.03},
during growth implies a non-zero component of their growth velocity in the horizontal direction. This leads to 
an observed motion of the fingers across the width of the simulation box (traveling waves of fingers) during eutectic 
colony growth.  

Emergence of a stable finger spacing can be observed in Fig.~\ref{ds_0_fi_v_0.015_del_0.03}, which is
not observed in the isotropic case. In the system with lower anisotropy shown 
in Fig.~\ref{ds_0_fi_v_0.015_del_0.015}, the fingers do broaden and bifurcate, but not as frequently as
in Fig.~\ref{iso_2D_no_ds}, which suggests that with an increase in the magnitude of anisotropy the stability
of the solid-liquid interface is enhanced.  
 
Similar to the isotropic situation, the lamellae appear to be oriented orthogonally to the 
solidification envelope at the solid-liquid interface 
which gets modified inside the fingers due to interactions between lamellae approaching 
the finger axis from either side of the finger tip. 

\begin{figure}[!htbp]
 \centering
 \subfigure[]{\includegraphics[width=0.4\linewidth]{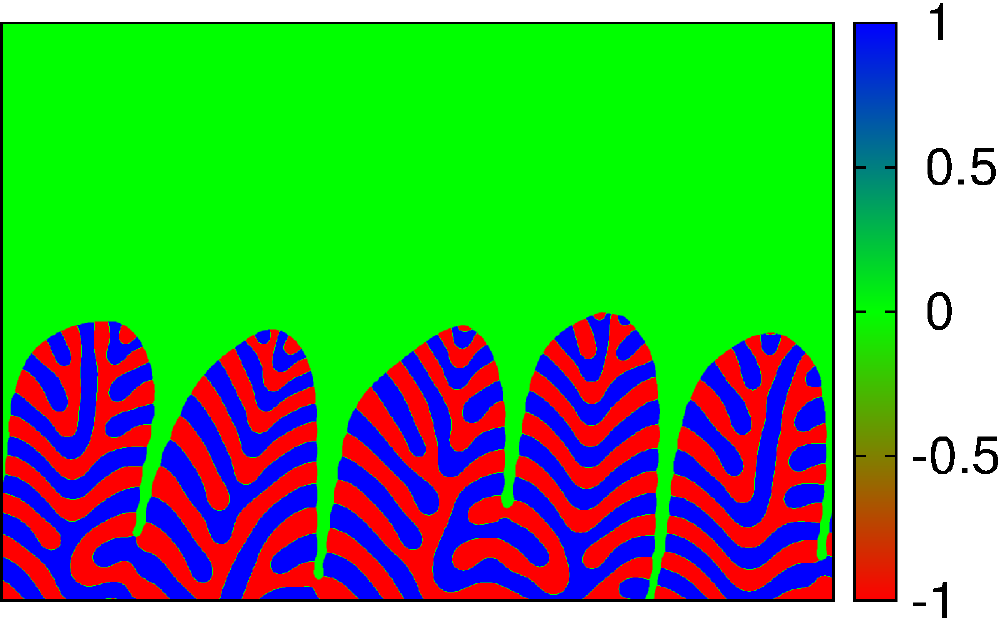}
 \label{ds_1_fi_v_0.04_del_0.03}
 }
 %\hspace{.25in}
 \subfigure[]{\includegraphics[width=0.4\linewidth]{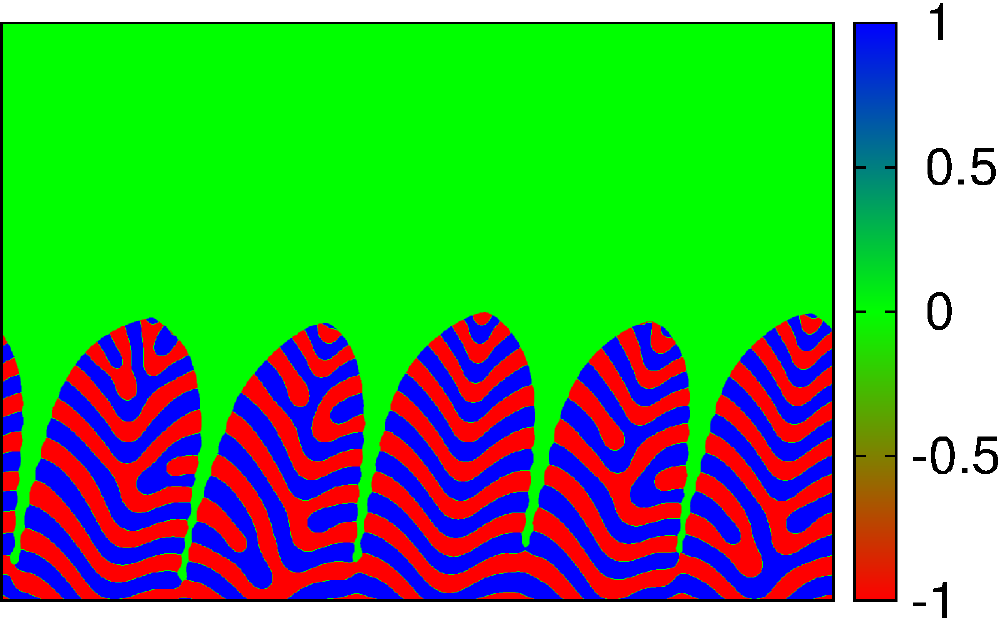}
 \label{ds_1_fi_v_0.04_del_0.05}
 }
%\subfigure[]{\includegraphics[width=0.3\linewidth]{ds_0_iso_v_0015_150000_c}
%\label{iso_2D_c_no_ds}
 %}
 %\centering
\caption{Microstructures ($u$ field) of a system having equal diffusivity in the solid and the liquid phases 
with solid-liquid interfacial energy anisotropy, for (a)$\delta=0.03$, $t=150000$, 
and (b)$\delta=0.05$, $t=100000$ with $V=0.04$ and $\theta_R=10\degree$, clockwise. Colorbars report values of the $u$ field. The other 
simulation parameters are the same as mentioned in the caption to Fig.~\ref{iso_2D_no_ds}.}
\label{ds_1_fi_aniso}
\end{figure}

An incomplete partitioning of the solutes (also known as solute trapping)
hindering the formation of the 
eutectic at the solid-liquid interface, is observed for
higher values of $V$ and $\delta$ when using solute mobilities of the form, 
$M=D\left(1-\phi^n\right)$ and $\widetilde{M}=\widetilde{D}\left(1-\phi^n\right)\tilde{c}$, even with higher values of $n$.
Thus, for studying the effects of higher $V$ and $\delta$ on the colony dynamics, we introduce 
another approximation in the form of equal and constant mobility of solutes in 
both solid and liquid phases. Mathematically, this manifests as setting $M=\widetilde{M}=1$ in the model formulation.
We report two simulations in Fig.~\ref{ds_1_fi_aniso} with $\delta=0.03$ (Fig.~\ref{ds_1_fi_v_0.04_del_0.03}) and 
$\delta=0.05$ (Fig.~\ref{ds_1_fi_v_0.04_del_0.05}) at a sample pulling velocity of $V=0.04$. 
The enhanced solid diffusivity should lead to a $\lambda_{JH}$ larger than the value computed for the system with no solid diffusivity
which should prevent length scales due to spinodal decomposition becoming dominant even at higher pulling velocities of $V=0.04$.

The fingers in Fig.~\ref{ds_1_fi_aniso} appear to possess a smaller tip-radius 
than what is observed for the cells in Fig.~\ref{ds_0_fi_aniso},
which is expected at higher velocities due to the inverse scaling of tip-radius ($\rho$), 
with pulling velocity ($V$) given by the constancy of $\rho^2 V$. Furthermore, an increase in $\delta$ can also be seen  
to promote a stronger selection of the dendrite tip radius as observed from Figs.~\ref{ds_1_fi_v_0.04_del_0.03} and~\ref{ds_1_fi_v_0.04_del_0.05}.  
Also, the tilt of
the fingers becomes more pronounced with increase in $\delta$, as can be confirmed from 
Figs.~\ref{ds_1_fi_v_0.04_del_0.03} and~\ref{ds_1_fi_v_0.04_del_0.05}. 

An interesting difference in the lamellar appearance can be observed at the central axis of the fingers.
In Fig.~\ref{ds_0_fi_aniso}, the individual phases from the solid-liquid interface on either side of the 
tip of a finger do not unite at the central axis of the finger, as they do in Fig.~\ref{ds_1_fi_aniso}.  
The tree-like arrangement of phases seen in Fig.~\ref{ds_0_fi_aniso} 
is not replicated in Fig.~\ref{ds_1_fi_aniso}, where the phases
from either side of the finger, join with each other in the middle of the fingers. This difference is a consequence of
the lack of solute diffusivity in the solid in Fig.~\ref{ds_0_fi_aniso}, where the local orthogonality of the lamellae to the
solidification envelope remains frozen even inside the fingers, with the emanation of phases from the central stem being a 
record of the lamellar bifurcation that has happened earlier.

Having considered the effect of anisotropic solid-liquid interfaces on the colony dynamics, we move on to studying systems 
with anisotropic solid-solid interfacial energies. 

\subsection{Anisotropic solid-solid interface}
In this section we are going to study the effect of an anisotropic interface between the two eutectic solids on the lamellar morphologies
constituting the eutectic colonies and also on the orientation and stability of the fingers.
We begin with a description of the phase-field model.   

\subsubsection{Phase-field model}
In order to explore the eutectic colony formation dynamics in situations 
where the solid-solid (i.e., $\alpha$-$\beta$) interfaces 
have a specific orientation with respect to the pulling direction, the anisotropy 
must be introduced through the $u$ field. But considering the fact $u$ 
changes in value across all the three possible interfaces 
($\alpha$-$\beta$, $\alpha$-liquid and $\beta$-liquid), 
we introduce the anisotropy through the 
the bulk free energy density in solid, to minimize its influence on 
the solid-liquid interface which results in the modified free-energy 
density expression of the solid given by,

\begin{equation}
f_{sol}=\frac{1}{8}(u^{2}-1)^{2} a_{c}^{2}(\theta) +  (\tilde{c} \ln \tilde{c} - \tilde{c} ) 
- (\ln K) \tilde{c} - \frac{\Delta T}{T_{E}},
\label{aniso_u_fsol}
\end{equation}
where $a_c$ is the same as in Eq.~\ref{ac_aniso_fi} with $\phi$'s being 
replaced by $u$'s. From Eq.~\ref{aniso_u_fsol}, we can see that the free 
energy density contribution from the $u$ field has a maximum at $u=0$ and 
minima at $u=\pm{1}$. Now by observing that the 
total energy density of the system is an interpolation between the 
solid and liquid energy densities through an interpolant($h$) which is a 
non-linear but monotonic function of $\phi$ (see Eq.~\ref{ftotal}), 
it can be verified that at a solid-liquid interface
($u$ varying between $0$ and $\pm{1}$; $\phi$ 
varying between $0$ and $1$) the influence of anisotropy is mellowed 
down by $u$ being non-zero and $h$ being non-unity.          
To confirm this observation, we can refer to Fig.~10 in~\cite{Lahiri2015}.
The equilibrium orientations of the $\alpha$-$\beta$ interfaces under solid-solid 
anisotropy can be 
discerned from Fig.~8 in~\cite{Lahiri2015}.

\subsubsection{Results}
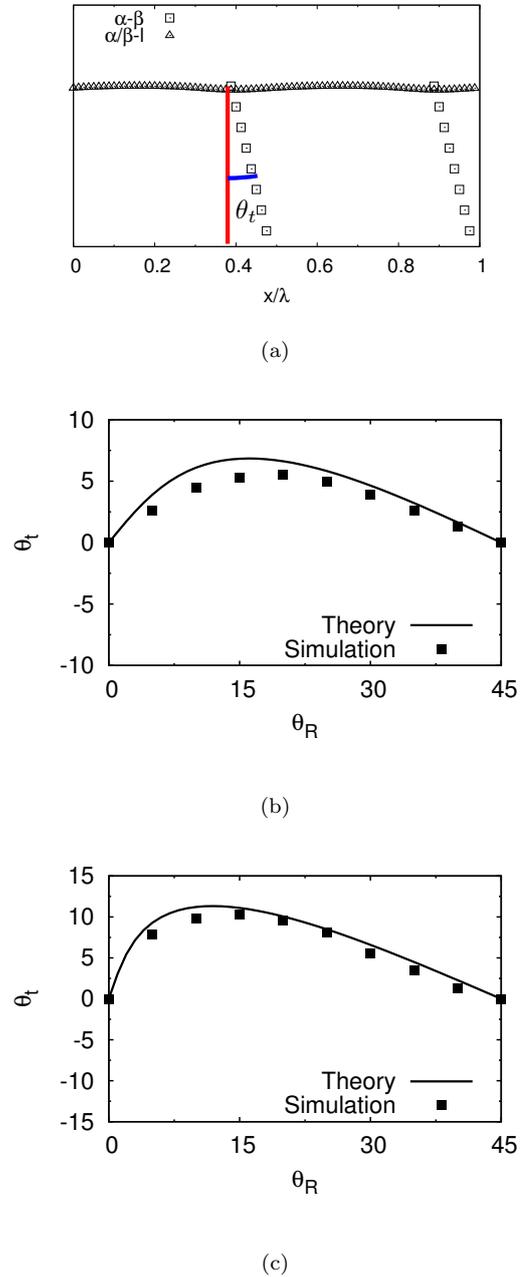
\begin{figure}[!htbp]
 \centering
 \subfigure[]{
 \begin{tikzpicture}
  \node[anchor=south west,inner sep=0] (image) at (0,0){\includegraphics[width=0.7\linewidth]{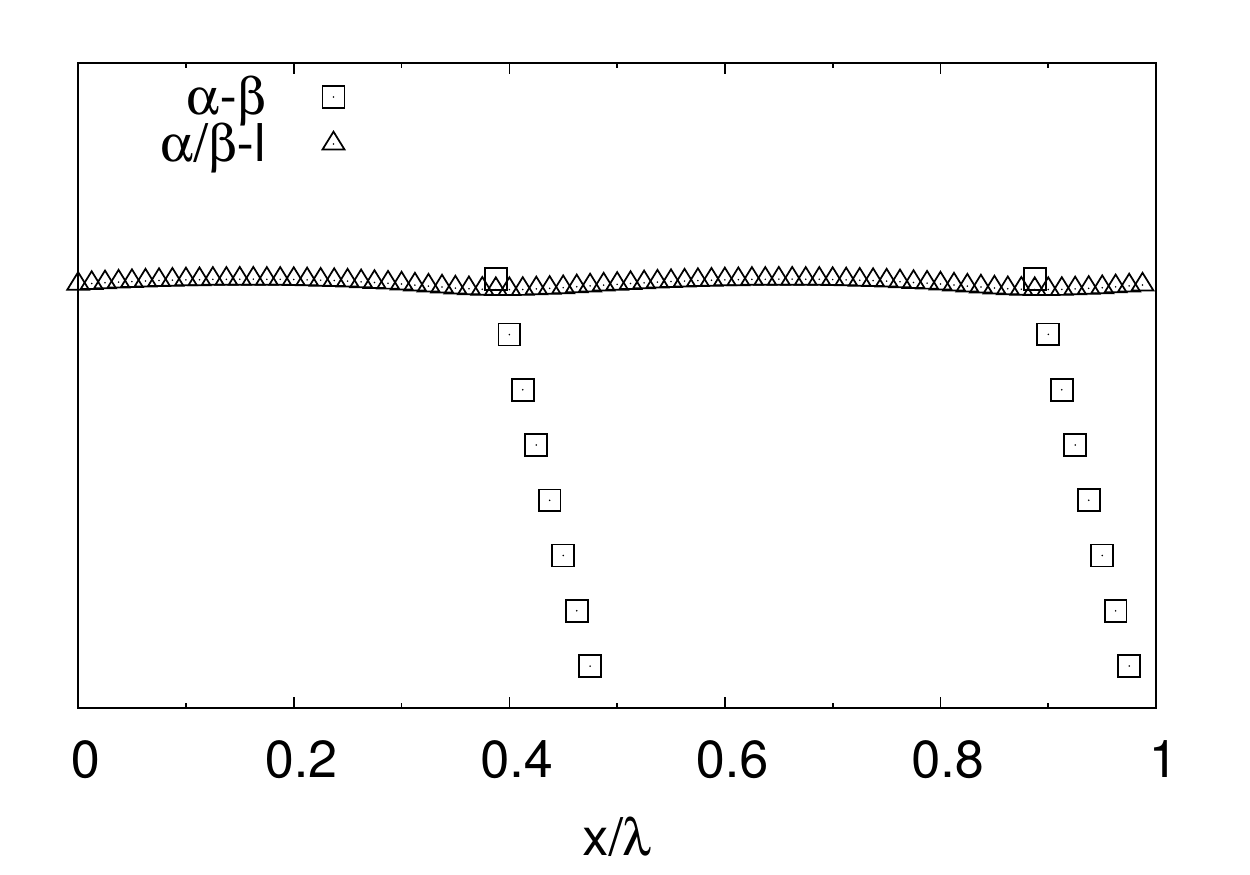}};
  \begin{scope}[x={(image.south east)},y={(image.north west)}]
   \draw [-,ultra thick,red] (0.39,0.2) --(0.39,0.68);
   \draw  [-,ultra thick,blue] (0.39,0.4) arc (-90:-77:0.28) ;
   \node[] at (0.43,0.3) {$\theta_t$};
   \end{scope}
   \end{tikzpicture}
 \label{interf}
 }
 \subfigure[]{\includegraphics[width=0.8\linewidth]{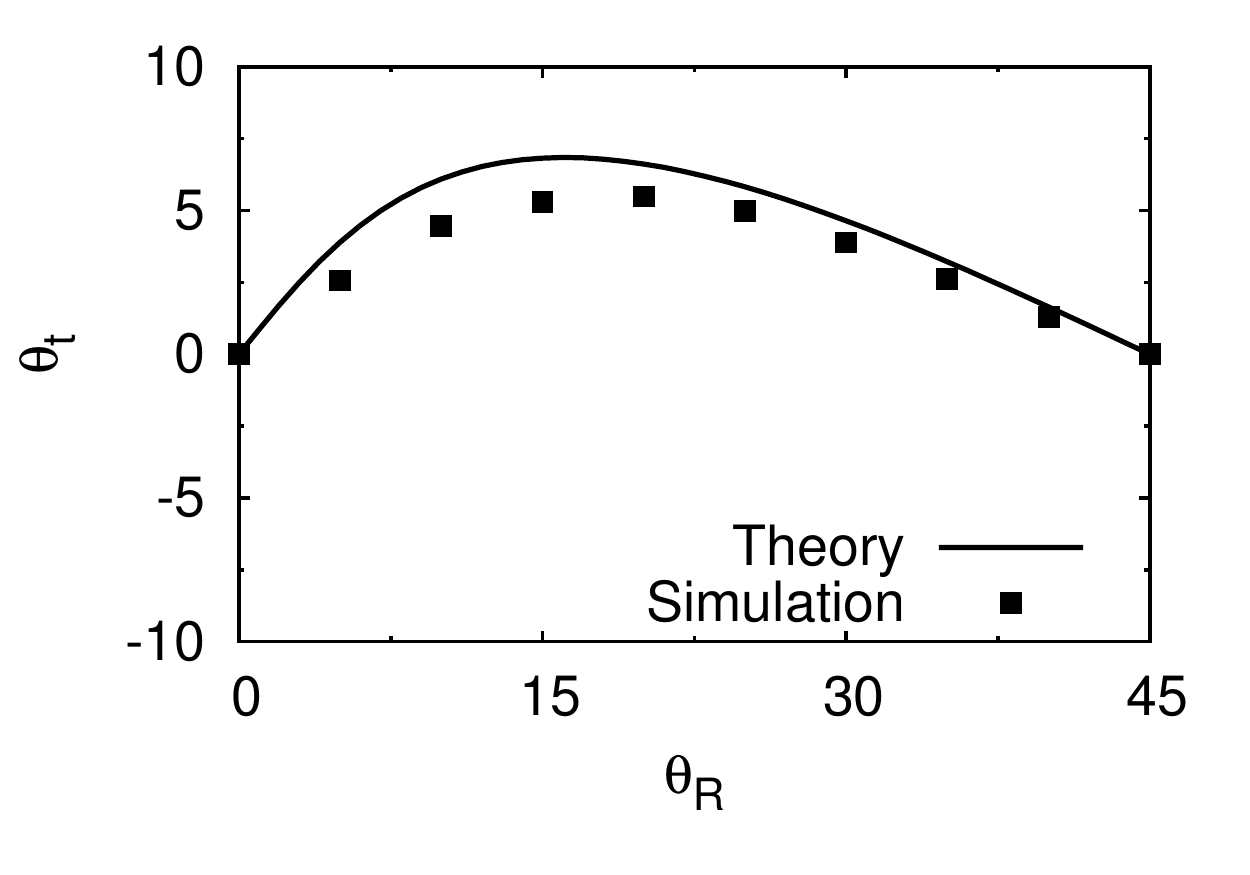}
 \label{tiltds0}
 }
 %\hspace{.25in}
 \subfigure[]{\includegraphics[width=0.8\linewidth]{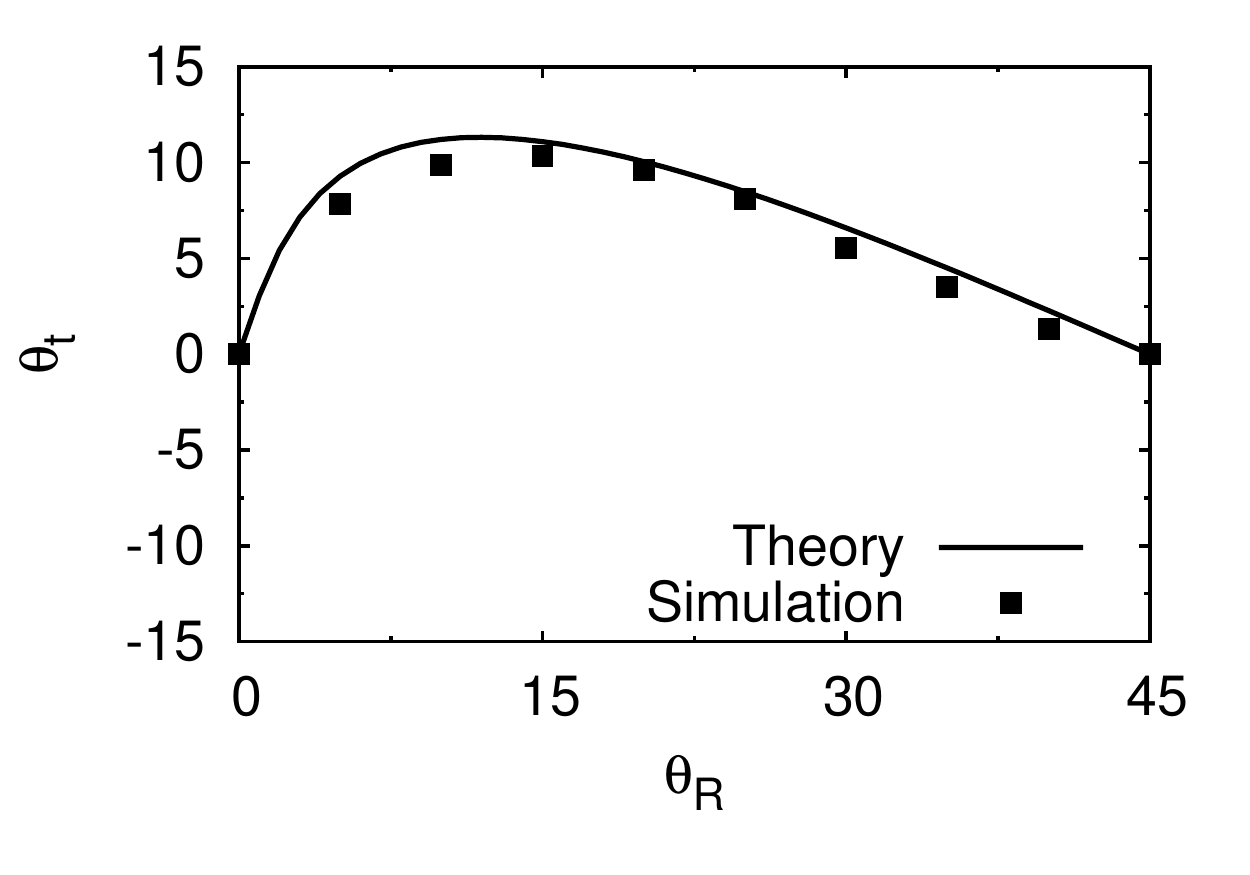}
 \label{tiltds1}
 }
\caption{Tilt of the solid-solid interface for $\theta_R=10\degree$, clockwise, 
and the corresponding tilt angle $\theta_t$ is shown in (a). 
Tilt angles ($\theta_t$) are plotted versus angle between the crystal frame and 
the laboratory frame ($\theta_R$), for (b) $M=\widetilde{M}=0$, for the eutectic solids; $\delta=0.03$, and
(c) $M=\widetilde{M}=1$ for the eutectic solids; $\delta=0.05$.}
\label{tilt_ds_0_1}
\end{figure}

For a given rotation of the crystal frame relative to the laboratory frame, 
the orientation of the $\alpha$-$\beta$ interface is going to
be determined by the force balance at the triple points. This can be predicted 
for the situation of steady-state growth using symmetry arguments that are motivated 
from experiments ~\cite{Akamatsu2012,Akamatsu2012_2} which claim that the 
resultant surface tension must still be oriented along the growth 
direction. This condition can then be used to derive 
an analytical expression for the $\alpha$-$\beta$ interfacial 
orientation with the vertical~\cite{Ghosh2015} which we are going to henceforth 
refer to as the tilt angle ($\theta_t$) (explained in Fig.~\ref{interf}).
The tilt of the solid-solid interface for 
a given rotation of the crystal frame can be seen in Fig.~\ref{interf}. For the four-fold anisotropy
function we have implemented, we compared the tilt angles from steady-state growth 
simulations of a single lamella pair against theoretical predictions in Fig.~\ref{tiltds0}.

\begin{figure}[!htbp]
 \centering
 \subfigure[]{\includegraphics[width=0.4\linewidth]{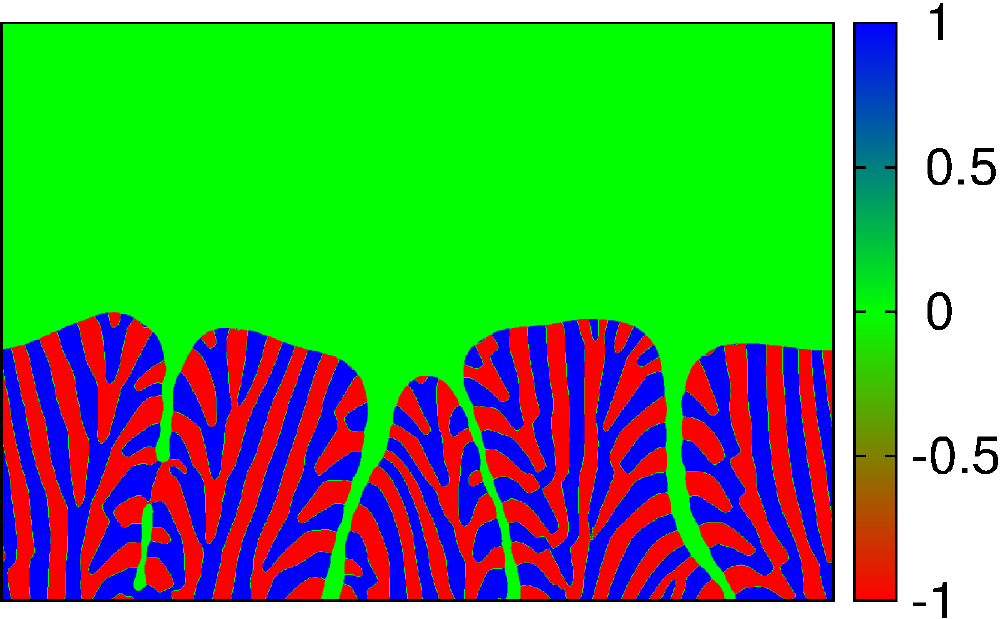}
 \label{ds_0_u_v_0.015_del_0.015_125000}
 }
 %\hspace{.25in}
 \subfigure[]{\includegraphics[width=0.4\linewidth]{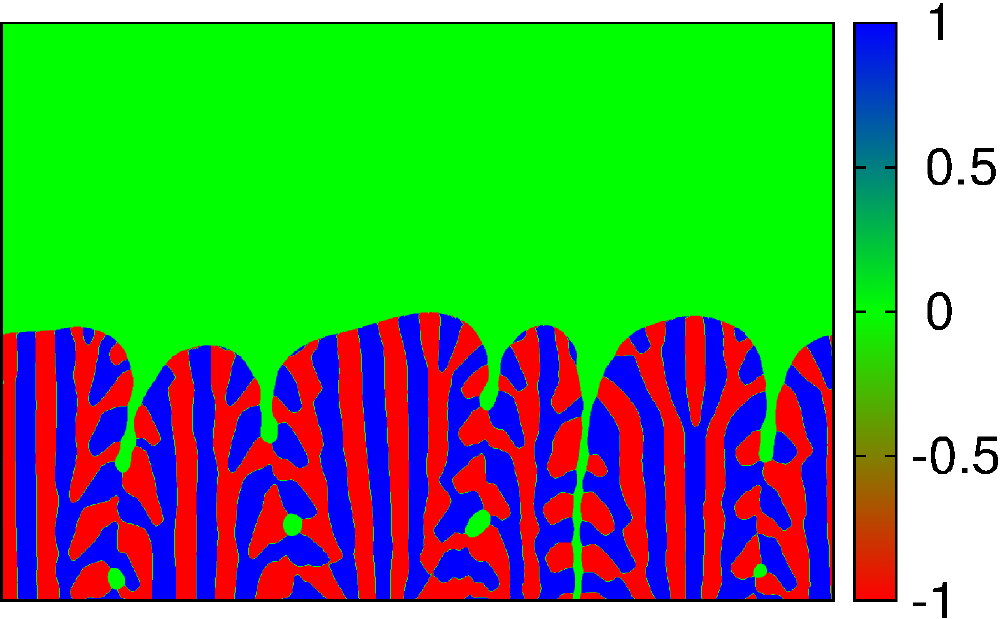}
 \label{ds_0_u_v_0.015_del_0.03_150000}
 }
%\subfigure[]{\includegraphics[width=0.3\linewidth]{ds_0_iso_v_0015_150000_c}
%\label{iso_2D_c_no_ds}
 %}
 %\centering
\caption{Microstructures ($u$ field) of a system with no solid diffusivity 
and solid-solid interfacial energy anisotropy, for (a)$\delta=0.015$, $n=16$, $t=125000$, 
and (b)$\delta=0.03$, $n=32$, $t=150000$, with $V=0.015$. Colorbars report values of the $u$ field. 
The other simulation parameters are the same as mentioned in the caption to Fig.~\ref{iso_2D_no_ds}.}
\label{ds_0_u_aniso}
\end{figure}

The colony formation dynamics with anisotropic solid-solid interfacial energy is explored in Fig.~\ref{ds_0_u_aniso} with two 
different magnitudes of anisotropy, i.e., $\delta=0.015$ in Fig.~\ref{ds_0_u_v_0.015_del_0.015_125000} and $\delta=0.03$
in Fig.~\ref{ds_0_u_v_0.015_del_0.03_150000} at a single pulling velocity of $V=0.015$.

Here, the microstructural feature that is strikingly different from the previous cases of isotropy and solid-liquid anisotropy
is the presence of straight, parallel, lamellae pairs running through the center of the fingers and are 
very similar to what is observed in Fig.~\ref{exp1}. This qualitative agreement between our simulations and experiments
substantiates our conjecture that the structures observed in Fig.~\ref{exp1} are a result of anisotropic interfaces
and more specifically anisotropy of the interface between the two eutectic solids.

Though there is a theoretical prediction available for the orientation of the $\alpha$-$\beta$ interfaces during 
steady-state growth, for an unsteady situation of cellular or dendritic growth 
we can only attempt to qualitatively understand the 
lamellar orientations in the absence of an
analytical expression. Focusing closely on Figs.~\ref{ds_0_u_v_0.015_del_0.015_125000} and~\ref{ds_0_u_v_0.015_del_0.03_150000},
the simulation proceeds from 
the destabilization of a tilted state of the lamellae during steady-state growth (as seen in Fig.~\ref{interf}),
giving rise to cells which globally
have lamellae oriented along the direction of the temperature gradient. The solidification 
envelope corresponding to each cell however develops small tilts with respect to the growth 
direction, which can be thought of as the tilted steady state at lower velocities 
being rotated such that the solid-solid interfaces become aligned with temperature gradient, 
while the solid-liquid interface develops a tilt.

The magnitude of anisotropy also appears to play a role in lamellar orientations, as for a smaller $\delta(=0.015)$,  as seen in 
Fig.~\ref{ds_0_u_v_0.015_del_0.015_125000}, the straight lamellae pairs are not strictly aligned with the vertical,
whereas with $\delta=0.03$, the lamellae display a strong alignment with the imposed temperature gradient, as can be
seen from Fig.~\ref{ds_0_u_v_0.015_del_0.03_150000}.

An outcome of the presence of lamellae oriented as closely as possible to the direction of imposed temperature gradient is the
broadening of fingers, as can be clearly observed in Fig.~\ref{ds_0_u_v_0.015_del_0.03_150000}. The shapes of the individual
fingers can be understood as a result of a combined influence of the propensity of the lamellae to remain aligned with the 
direction of the imposed temperature gradient and that of the thermodynamically predicted relative orientations of the solid-solid and
the solid-liquid interfaces under solid-solid interfacial energy anisotropy presented in Fig.~\ref{tilt_ds_0_1}.
Thus, the vertically oriented lamellae emanate
from sides of the fingers which appear roughly flat, but display a small deviation from horizontal. 
% The requirements that the lamellae be 
% oriented vertically while obeying the force balance at the triple points under anisotropic solid-solid interfaces, favour broad, 
% slightly tilted, solid-liquid interfaces, which allows the lamellae to satisfy both conditions simultaeneously. 

Furthermore, with regards to the stability of the finger width, an increase in $\delta$ leads to stabler features, which is the same as seen for
the case of anisotropic solid-liquid interfacial energy.

\begin{figure}[!htbp]
 \centering
 \subfigure[]{\includegraphics[width=0.4\linewidth]{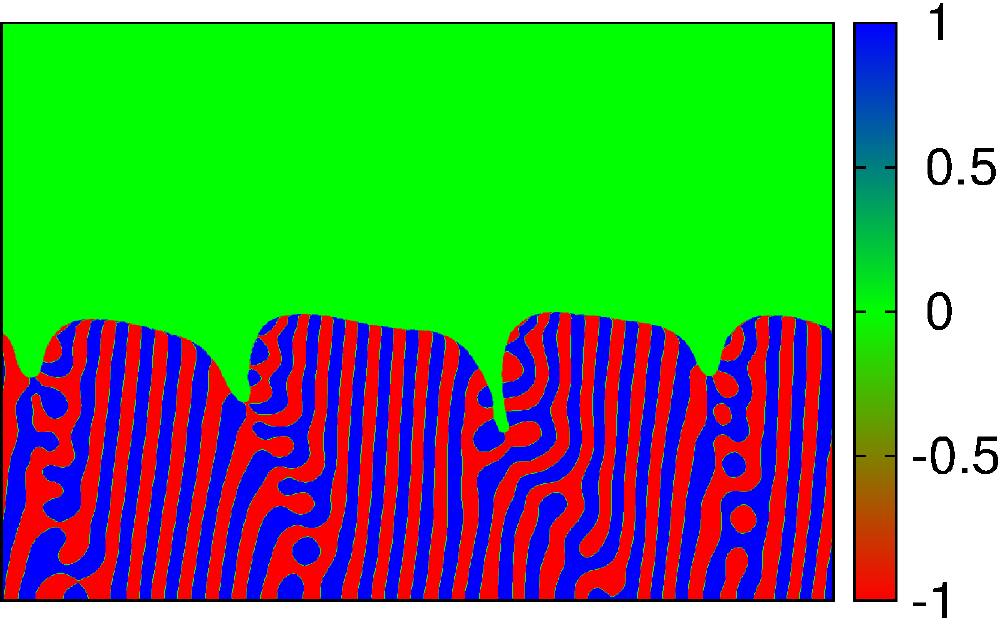}
 \label{ds_1_u_v_0.04_del_0.03_150000}
 }
 %\hspace{.25in}
 \subfigure[]{\includegraphics[width=0.4\linewidth]{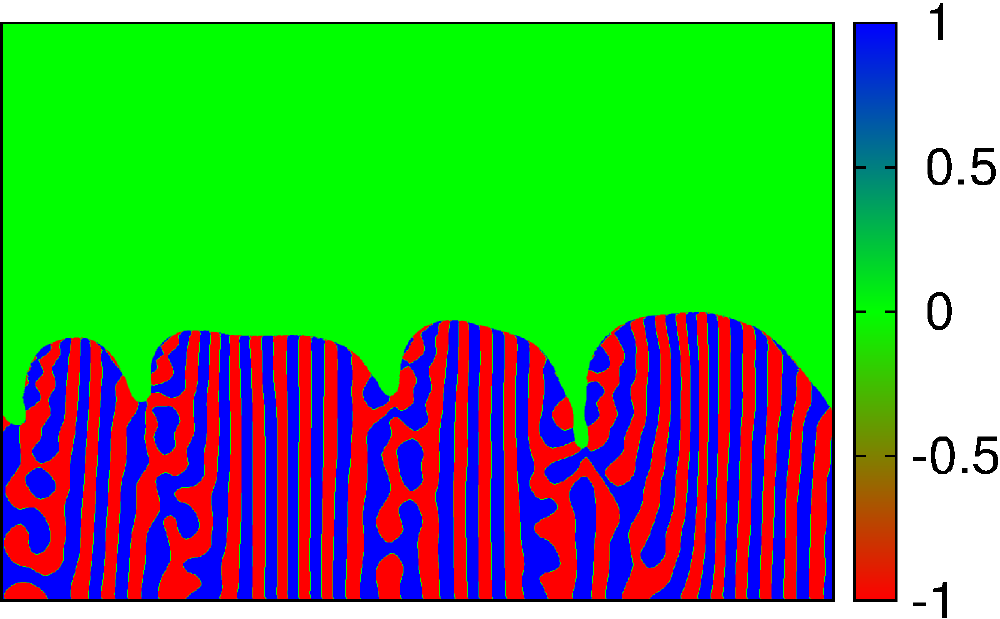}
 \label{ds_1_u_v_0.04_del_0.05_150000}
 }
%\subfigure[]{\includegraphics[width=0.3\linewidth]{ds_0_iso_v_0015_150000_c}
%\label{iso_2D_c_no_ds}
 %}
 %\centering
\caption{Microstructures ($u$ field) of a system with equal solute diffusivity in the 
solid and the liquid phases, with solid-solid interfacial energy anisotropy, for (a)$\delta=0.03$; $t=150000$ 
and (b)$\delta=0.05$; $t=150000$, with $V=0.04$. Colorbars report values of the $u$ field. 
The other simulation parameters are the same as mentioned in the caption to Fig.~\ref{iso_2D_no_ds}.}
\label{ds_1_u_aniso}
\end{figure}

As mentioned in conjunction to the discussion on colony dynamics with solid-liquid anisotropy, we relax the criterion of negligible
solute diffusivities in the solid in order to negate the solute-trapping that impedes probing of 
colony formation in systems pulled at higher velocities or displaying higher anisotropy in the interfacial energy. A variation of tilt angles
($\theta_t$) as a function of $\theta_R$ for equal diffusivity of solutes in the solid and the liquid is also reported in Fig.~\ref{tiltds1} 
as a confirmation that this approximation preserves the ability of the model to capture the essential physics.  
This enables us to simulate the effects of a pulling velocity of $V=0.04$ on systems
with $\delta=0.03$ (see Fig.~\ref{ds_1_u_v_0.04_del_0.03_150000}) and $\delta=0.05$ (see Fig.~\ref{ds_1_u_v_0.04_del_0.05_150000}). 

Most of the features seen in Fig.~\ref{ds_0_u_aniso} is replicated in Fig.~\ref{ds_1_u_aniso} 
except for a few exceptions. One of them being the absence 
of the deep cells observed in Fig.~\ref{ds_0_u_aniso}. This is a result of the enhanced diffusivity in the 
solid which allows adjoining fingers to fuse wherever they are in close proximity.
Another important feature of these simulations is the lateral orientation (towards the left in Fig.~\ref{ds_1_u_aniso}) 
of the two-phase finger-tips in their bid to choose a smaller tip radius ($\rho$) consistent with a larger pulling velocity $V$ while 
allowing the maximum number of lamellae to remain vertical at the same time maintaining the necessary orientation relationship
between the solid-solid and solid-liquid interfaces.
Furthermore, the lamellae in Fig.~\ref{ds_1_u_aniso} also appears finer consistent with the
higher pulling velocities employed for these simulations~\footnote{$\lambda$ scales as $\rho$ with change in velocity, 
with the scaling constant depending on the simulation conditions. Thus, changes in lamellar widths can be understood in the context
of the concurrent changes tip radius with velocity ($V$), which is also indicative of the magnitude of the scaling constant connecting
$\lambda$ and $\rho$ for the current simulation.}.

% Thus to summarize, the principal observation from the colony morphologies obtained in a situation with anisotropic interfaces
% between the two eutectic solids is the presence of lamellae which are oriented along the imposed temperature gradient. This also 
% modifies the shapes of the fingers which can be understood on the basis of the orientational relationship between the solid-solid
% and the solid-liquid interfaces. The stability of a finger width was found to improve with increase 

Having studied the lamellar orientations and the two-phase cell morphologies for systems with anisotropic solid-solid and solid-liquid
interfacial energies in 2D, we now move on to 3D simulations where we probe the effect of a third dimension on the colony dynamics
in systems with anisotropic interfacial energies. We begin with a discussion of an isotropic system.

\section{3D: Isotropic}
The 2D simulations provide important insights into the physics of the colony formation problem in terms of both lamellar
and finger morphologies. But these observations from 2D simulations suffer from a limitation of being representative only of
directional solidification in thin samples. In order to gain a complete understanding of the problem in situations where both dimensions
of the solidified cross-section are comparable, we resort to 3D studies beginning with the isotropic system. The 
governing equations \ref{fi_evol},~\ref{u_evol} and~\ref{c_evol} are expressed in a tensorial form which  
are numerically solved in a 3D cartesian system. 

The high computational cost of 3D simulations, constrain us to perform them for 
a set of parameters which lead to a quicker destabilization of the solidification front.
Hence, we employ a high pulling velocity of $V=0.1$ to computationally access the colony dynamics
in 3D. This approach necessitates equal diffusivity of solutes in the solid to that in the liquid,
due to the high solute trapping observed in this model at higher pulling speeds.

It must be mentioned at this point that though for $V=0.1$ the spinodal decomposition length 
scale $\lambda_{max}^{spin}$ becomes larger than $\lambda_{JH}$, the scale of the simulation 
in 3D is always maintained at $\lambda_{JH}$ due to the presence of 3D topological mechanisms 
for lamellar interactions.

\begin{figure}[!htbp]
\centering
%\begin{tikzpicture}[scale=1]
%\draw (5,-0.1) --(7,-2);
\subfigure[]{\includegraphics[width=0.3\linewidth]{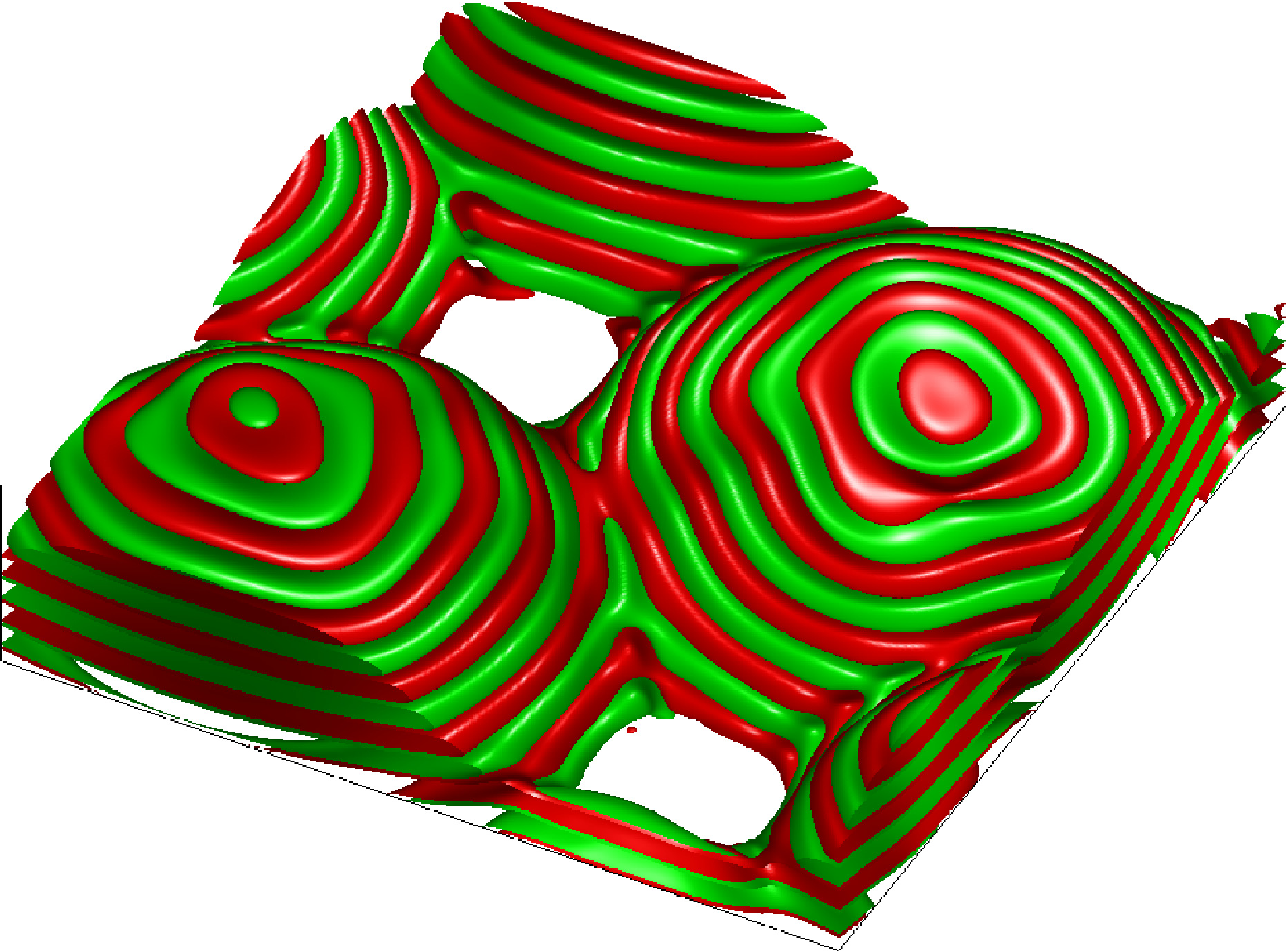}
\label{iso_3D_1}
}
\subfigure[]{\includegraphics[width=0.3\linewidth]{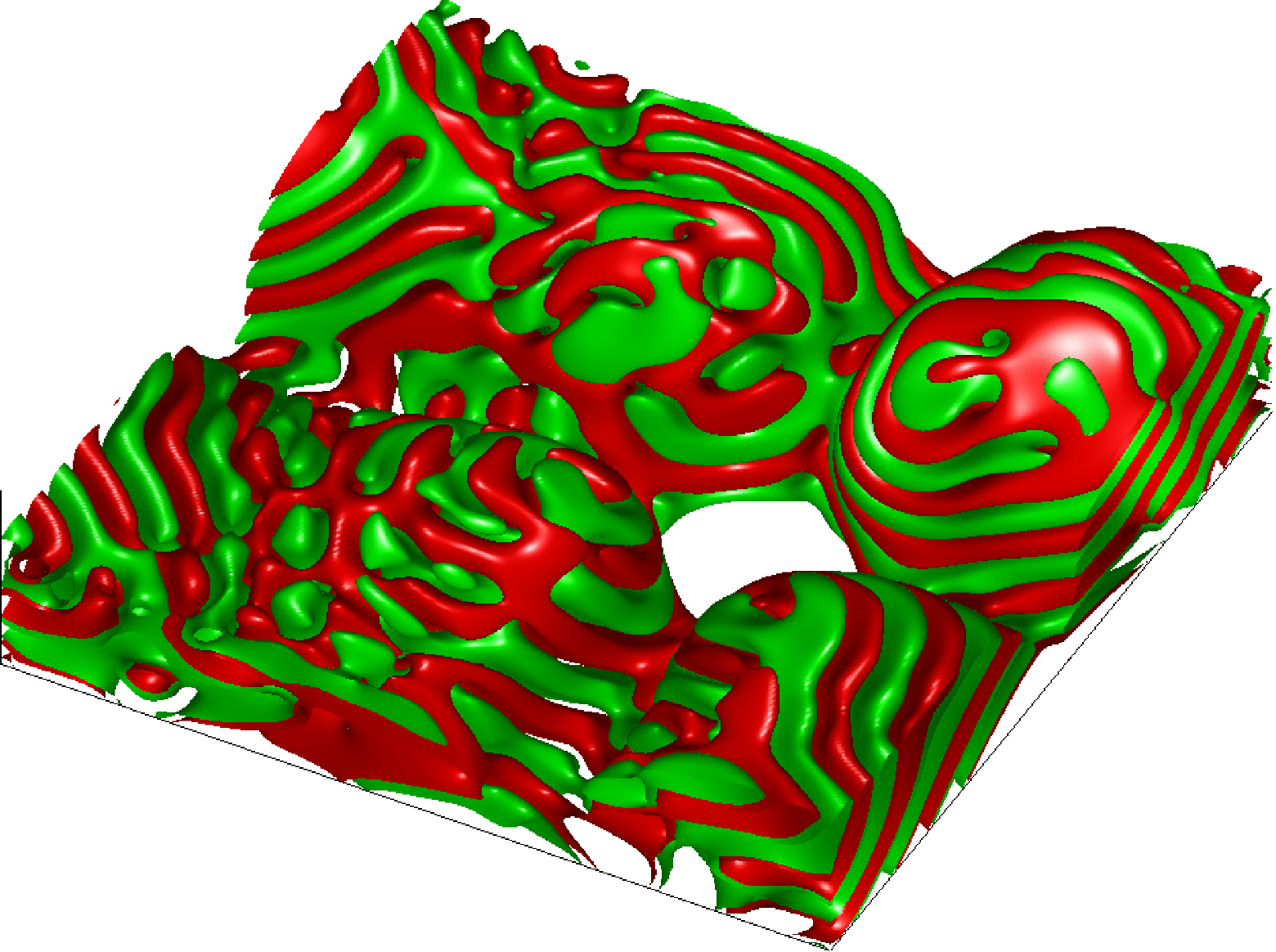}
\label{iso_3D_2}
}
%\hspace{.25in}
%\subfigure[]{\includegraphics[width=0.3\linewidth]{isotropic_55000_3D}
\subfigure[]{
\begin{tikzpicture}
  \node[anchor=south west,inner sep=0] (image) at (0,0){\includegraphics[width=0.3\linewidth]{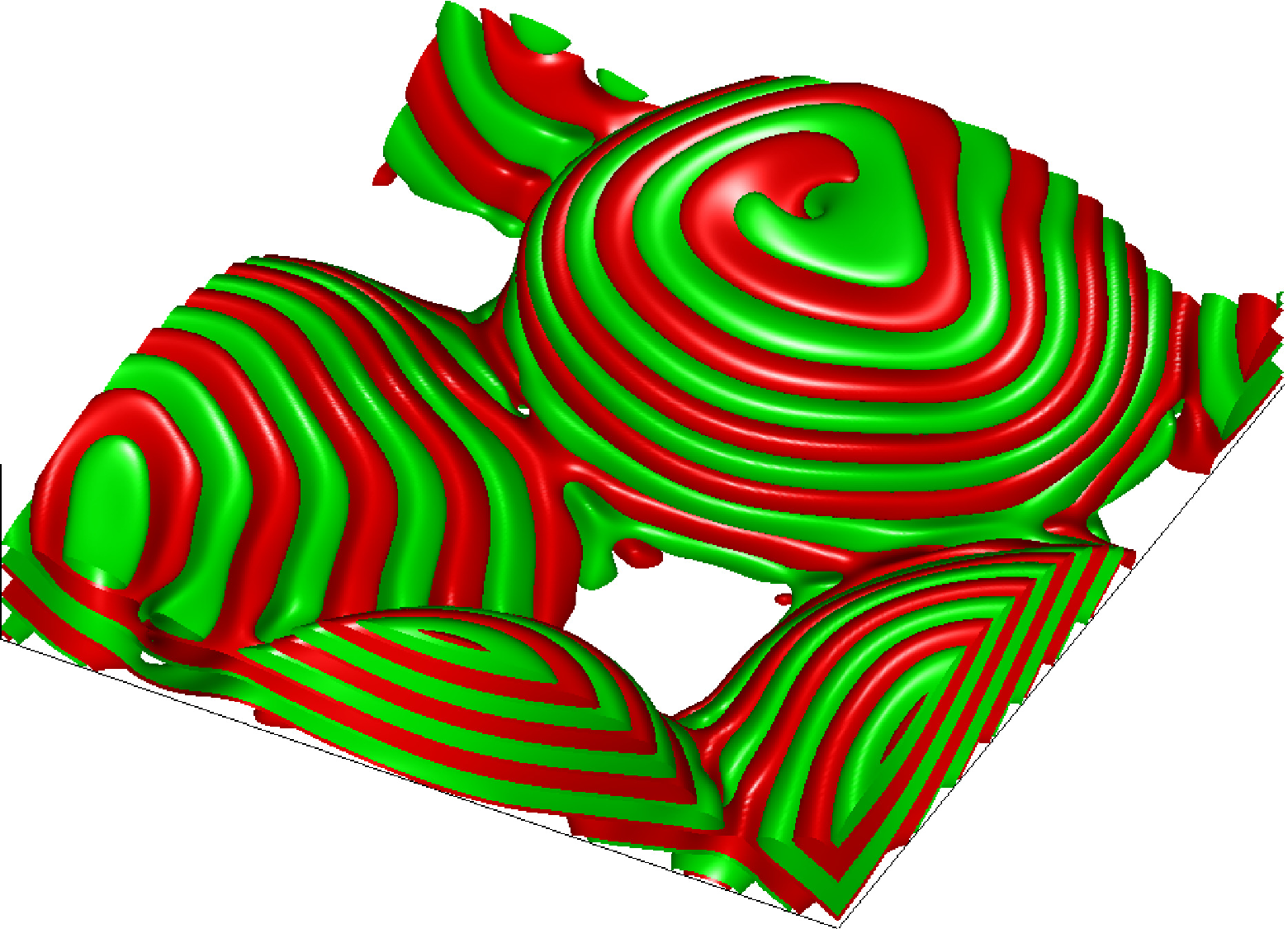}};
  \begin{scope}[x={(image.south east)},y={(image.north west)}]
   \draw [dashed,ultra thick,yellow] (0.3,0.9) --(1.0,0.5);
   \end{scope}
   \end{tikzpicture}
\label{iso_3D_3}
}
%\end{tikzpicture}
%\centering
\caption{Eutectic colonies at a total time of (a)21000, (b)45000, and (c)55000 in an isotropic system with $V=0.1$. The yellow dashed line in 
(c) represents the orientation of the vertical sections reported in Fig.~\ref{iso_fi_sect_y}. }
\label{iso_3D}
\end{figure}

We report a 3D simulation in Fig.~\ref{iso_3D} carried out in 
a $288\times288\times300$ box 
containing $8$ lamellae pairs along each dimension 
with the remaining parameters being identical to the 2D isotropic 
simulation shown in Fig.~\ref{ds_1_fi_aniso}. 
The simulations are done in a directional solidification setting with the 
direction of the imposed temperature gradient being vertically upwards.
The boundary conditions are set to be no-flux on faces of the box normal to the 
pulling direction and periodic on faces parallel to it. 

In Fig.~\ref{iso_3D_3}, a single lamella pair appears to construct each finger by
growing continuously in a helical fashion. This structure has already been observed experimentally by
Akamatsu and Faivre~\cite{Akamatsu2010} and has been anointed by them as a ``spiraling eutectic''. 

\begin{figure}[!htbp]
\centering
\subfigure[]{\includegraphics[width=0.4\linewidth]{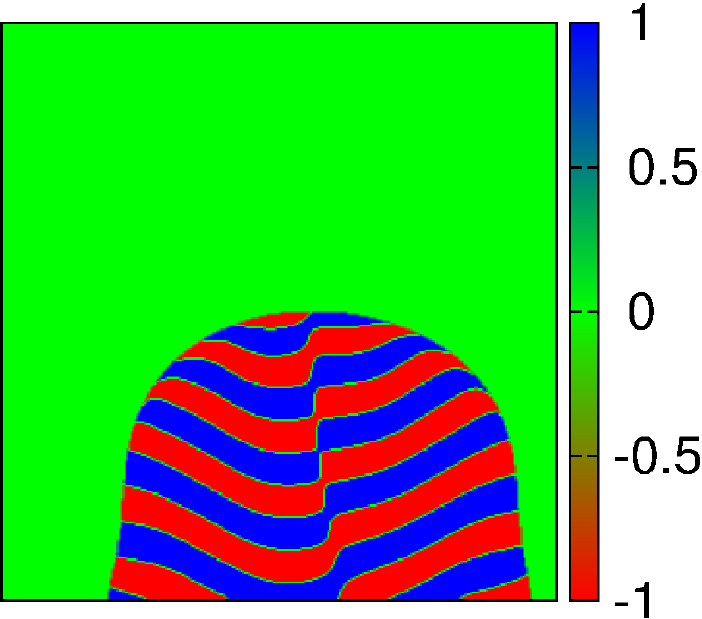}
\label{iso_fi_sect_y_1}
}
%\hspace{.25in}
\subfigure[]{\includegraphics[width=0.4\linewidth]{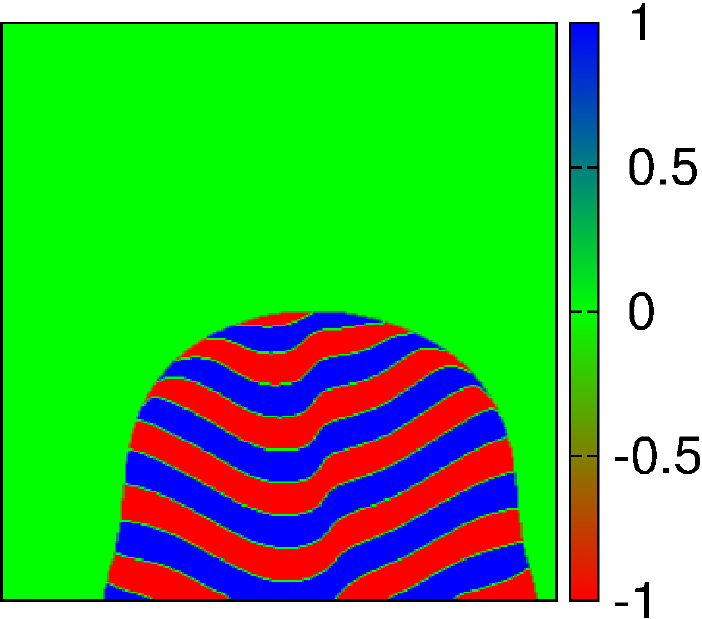}
\label{iso_fi_sect_y_2}
}
%\centering
\caption{2D sections ($u$ field) of Fig.~\ref{iso_3D_3} by a plane parallel to the pulling direction whose orientation is indicated 
by the yellow dotted
line in Fig.~\ref{iso_3D_3}, showing solid phase arrangements 
at (a)the spiral axis, and (b)a little away from the spiral axis. Colorbars report values from the $u$ field.}
\label{iso_fi_sect_y}
\end{figure}

\begin{figure}[!htbp]
\centering
\subfigure[]{\includegraphics[width=0.4\linewidth]{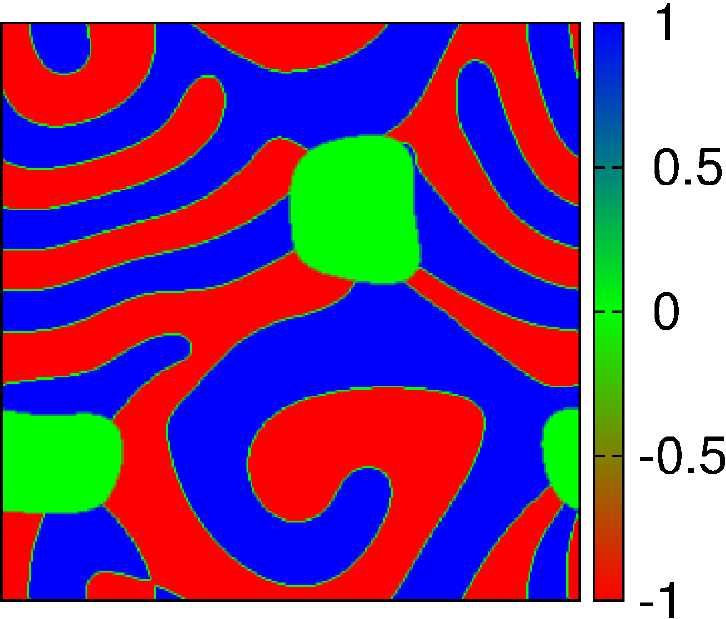}
\label{iso_fi_sect_z_1}
}
%\hspace{.25in}
\subfigure[]{\includegraphics[width=0.4\linewidth]{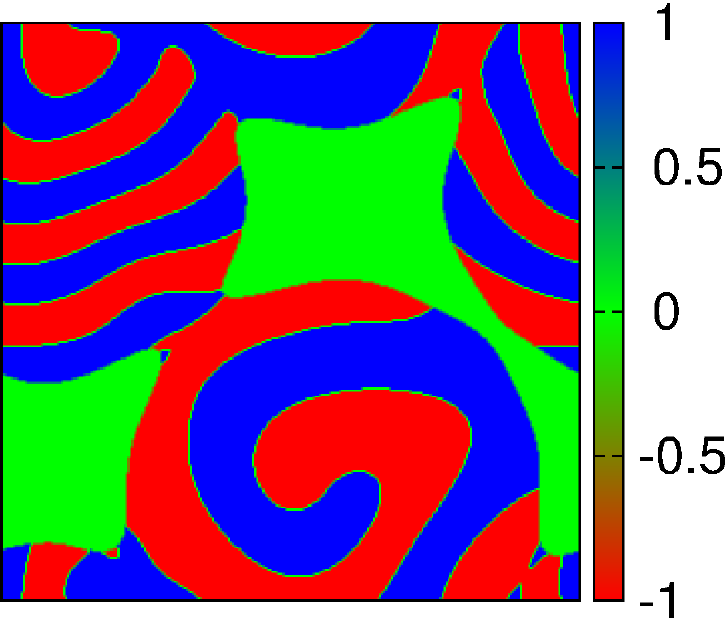}
\label{iso_fi_sect_z_2}
}
%\centering
\caption{2D sections ($u$ field) of Fig.~\ref{iso_3D_3} by a plane normal to the pulling direction.
The sectioning height is lower in (a) than in (b). Colorbars report values from the $u$ field.}
\label{iso_fi_sect_z}
\end{figure}

The morphology of the spiral can be better understood by considering a 2D sections of the microstructure in Fig.~\ref{iso_3D_3} 
by planes parallel and perpendicular to the growth direction as reported in Figs.~\ref{iso_fi_sect_y} and~\ref{iso_fi_sect_z}, respectively. 
The apparent discontinuity in the solid phases across the central axis of the finger seen in Fig.~\ref{iso_fi_sect_y_1} coupled with 
the particular arrangement of phases in Fig.~\ref{iso_fi_sect_z} allow an understanding of the spiral as a helical arrangement of 
a single pair of lamella plates.
Figs.~\ref{iso_3D_3},~\ref{iso_fi_sect_y} and~\ref{iso_fi_sect_z}
considered in unison points to the possibility of the shape of the spiraling 
eutectic fingers being approximated by a paraboloid.  
% This coupled with the information provided by Fig.~\ref{iso_fi_sect_y_2} lends to an understanding of the spiral morphology as one 
% where plates of individual phases ($\alpha$ or $\beta$) alternately occupy a single pitch of the helix whose axis is the same as that of    
% the spiral. This corrects and clarifies the impression from Fig.~\ref{iso_3D_3} that only a single helically advancing lamella pair 
% constitutes an eutectic spiral, as Fig.~\ref{iso_fi_sect_y} shows that a very specific arrangement of an array of lamellae pairs gives rise 
% to the spiral morphology. This morphology of the spiral for eutectic compositions is also confirmed by the observations in~\cite{Pusztai2013}.
% From the section of a finger in a plane perpendicular to the growth direction (see Fig.~\ref{iso_fi_sect_z}), 
% the increase in the radius of the individual 
% phase plates as they appear lower in the spiral can be discerned.
%  Furthermore, in Fig.~\ref{iso_fi_sect_z} the phases resemble annular arcs 
% with the center of curvature located inside the fingers away from the solid-liquid interfaces.

The 3D simulation of the isotropic system confirms the major observations from the 
2D simulation in terms of the randomness of the finger orientations,
the lack of specificity of the lamellar orientations and the absence of a particular finger 
spacing selected by the same. The solid-liquid interface is found to be unstable 
with spirals forming and disintegrating throughout the course of the simulation as can be seen 
by considering the Figs.~\ref{iso_3D_1},~\ref{iso_3D_2} and ~\ref{iso_3D_3}.

The discussion of an isotropic system in 3D, provides a reference against which we will seek to understand the 
eutectic colony features under interfacial anisotropy presented in the following sections. 

\section{3D: Effect of anisotropic interfacial energies on the colony dynamics}
Like in 2D, we will consider the effect of both solid-liquid and solid-solid anisotropy on the colony
formation dynamics in 3D. We begin with the former.

\subsection{Anisotropic solid-liquid interface}
To explore the effect of anisotropy on the solid-liquid interface on the 
colony formation dynamics in a 3D system, we introduce the anisotropy 
through the $\phi-$ field with the expression for $a_c$ being:

\begin{equation}
 a_c=
 1-\delta \left(3-4\left(\frac{{\phi^*}_{x}^{4}+{\phi^*}_{y}^{4}+{\phi^*}_{z}^{4}}{({\phi^*}_{x}^{2}+{\phi^*}_{y}^{2}+
 {\phi^*}_{z}^{2})^{2}}\right)\right),
\label{ac_fi_3D}
\end{equation}

which is a simple extension of the 2D case. In reality, 
the crystal frame can have 
any arbitrary orientation to the pulling direction. 
But any such
orientation can be decomposed into a combination of a rotation about 
the pulling direction and the ones normal to it. In view of 
that, we can attempt to understand the microstructure 
formation for two basic configurations: the one where the axis of 
rotation of the crystal frame is the same as the pulling 
direction and the other where it is perpendicular.
We will begin the discussion with the former.

\subsubsection{Crystal frame rotated about the pulling direction}
\begin{figure}[!htbp]
\centering
\subfigure[]{\includegraphics[width=0.4\linewidth]{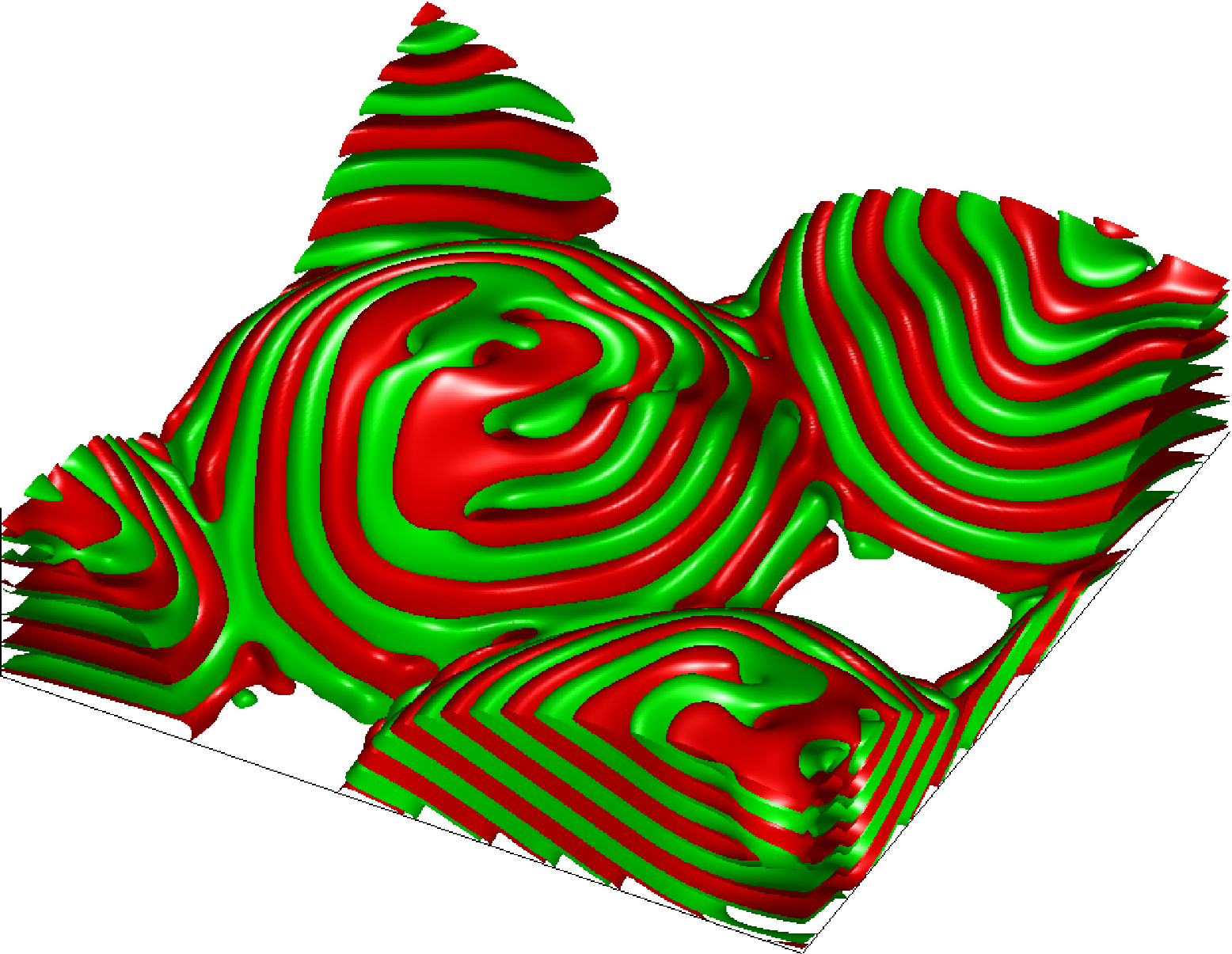}
\label{fi_z_3D_1}
}
%\subfigure[]{\includegraphics[width=0.4\linewidth]{fi_z_55000}
\subfigure[]{
\begin{tikzpicture}
  \node[anchor=south west,inner sep=0] (image) at (0,0){\includegraphics[width=0.4\linewidth]{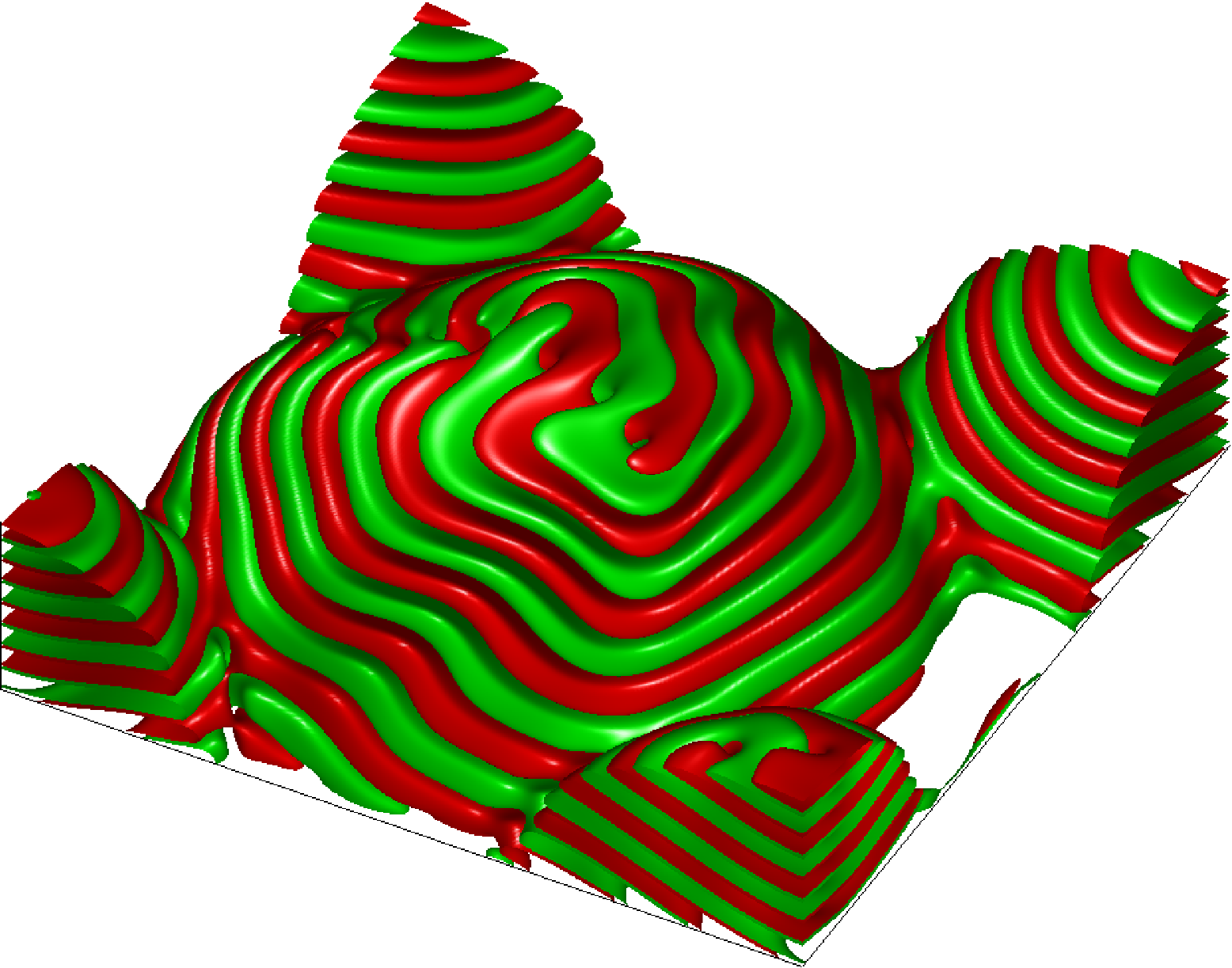}};
  \begin{scope}[x={(image.south east)},y={(image.north west)}]
   \draw [dashed,ultra thick,yellow] (0.25,0.15) --(0.65,0.9);
   \end{scope}
   \end{tikzpicture}
\label{fi_z_3D_2}
}
%\centering
\caption{Eutectic colonies at a total time of (a)28000, and (b)55000 in a system with 
solid-liquid interfacial energy anisotropy 
with $V=0.1$, $\delta=0.015$, and $\theta_R=10\degree$, clockwise, about the pulling direction. The yellow dashed line in 
(b) represents the orientation of the vertical sections reported in Fig.~\ref{fi_z_sect_x}. }
\label{fi_z_3D}
\end{figure}

\begin{figure}[!htbp]
\centering
\subfigure[]{\includegraphics[width=0.4\linewidth]{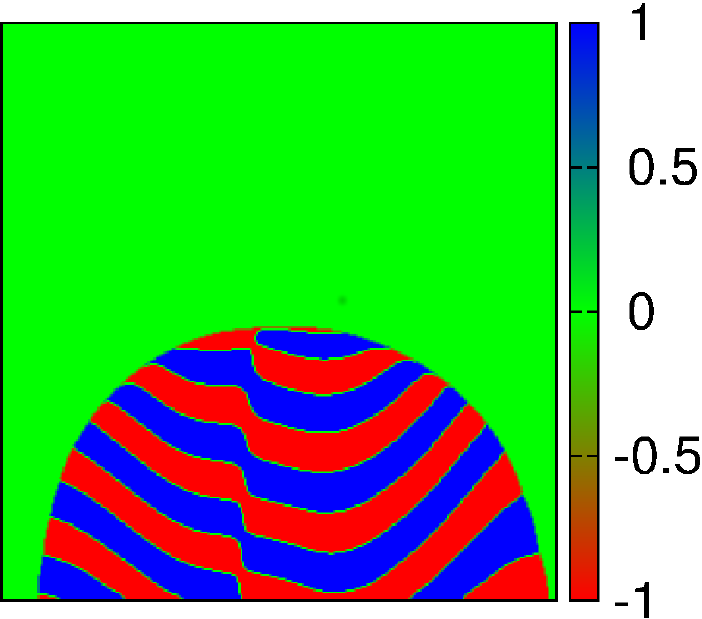}
\label{fi_z_sect_x_1}
}
%\hspace{.25in}
\subfigure[]{\includegraphics[width=0.4\linewidth]{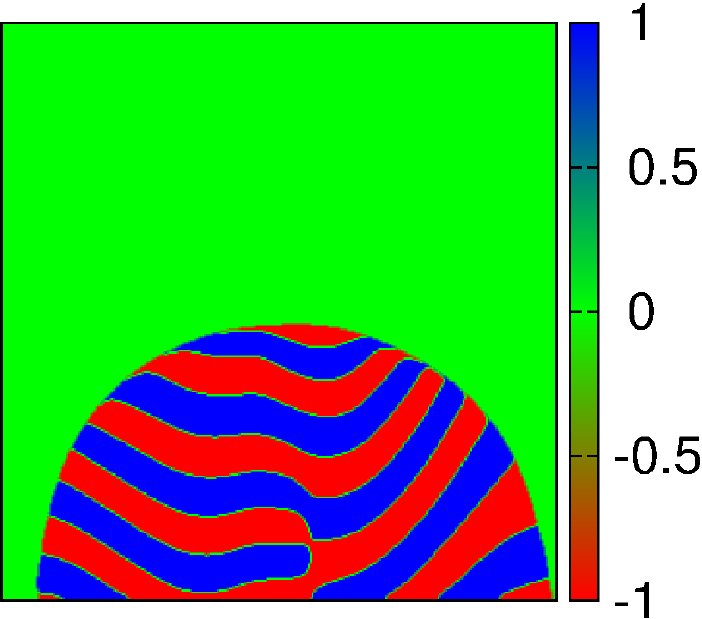}
\label{fi_z_sect_x_2}
}
%\centering
\caption{2D sections ($u$ field) by a plane parallel to the pulling direction in Fig.~\ref{fi_z_3D_2}, whose orientation is 
denoted by the yellow dotted line in Fig.~\ref{fi_z_3D_2}. The section passes through 
and a little away from the axis of the finger in (a) and (b), respectively. Colorbars report values from the $u$ field.}
\label{fi_z_sect_x}
\end{figure}

\begin{figure}[!htbp]
\centering
\subfigure[]{\includegraphics[width=0.4\linewidth]{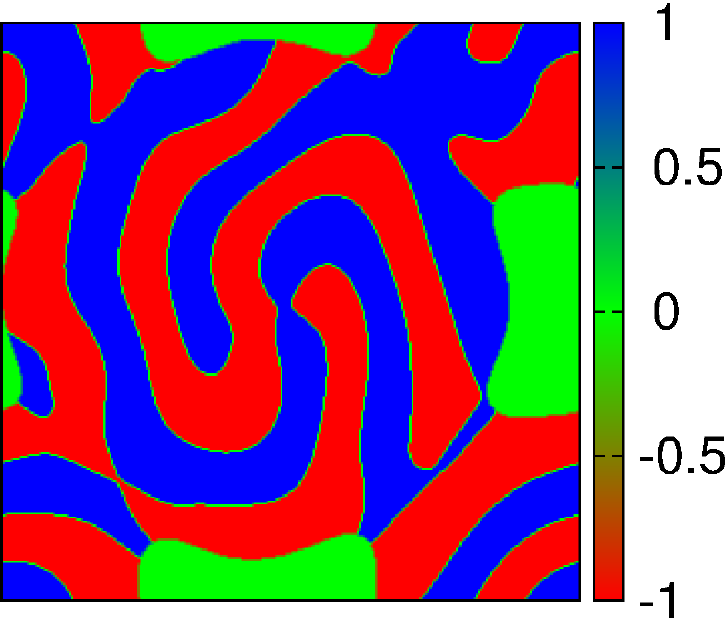}
\label{fi_z_sect_z_1}
}
%\hspace{.25in}
\subfigure[]{\includegraphics[width=0.4\linewidth]{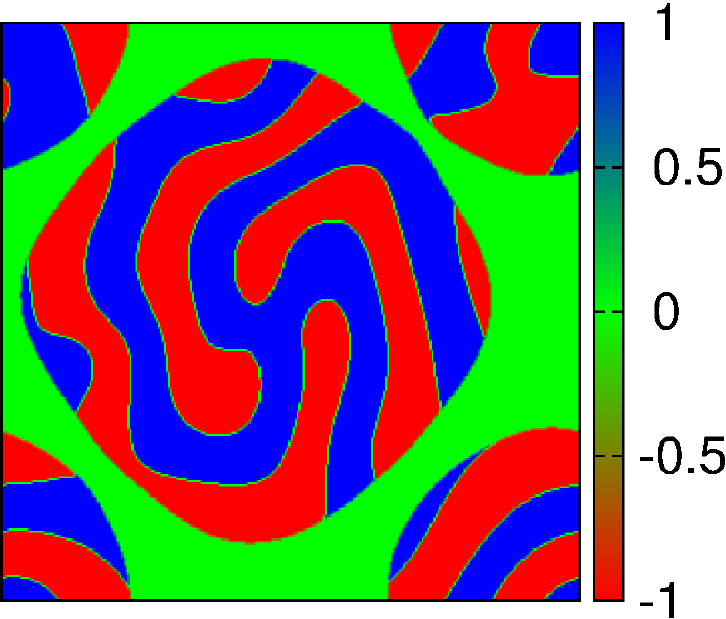}
\label{fi_z_sect_z_2}
}
%\centering
\caption{2D sections ($u$ field) by a plane normal to the pulling direction in Fig.~\ref{fi_z_3D_2}.
The sectioning height is lower in (a) than in (b). Colorbars report values from the $u$ field.}
\label{fi_z_sect_z}
\end{figure}

The microstructure shown in Fig.~\ref{fi_z_3D_2} is similar to Fig.~\ref{iso_3D_3} 
in terms of lamellar and finger morphologies with 
the influence of anisotropy showing up clearly only in the transverse section 
of the finger (see Fig.~\ref{fi_z_sect_z}) which becomes polygonal. The longitudinal 
section in Fig.~\ref{fi_z_sect_x} reveals a lamellar arrangement
which is also akin to its isotropic counterpart in Fig.~\ref{iso_fi_sect_y}.

\subsubsection{Crystal frame rotated about an axis normal to the pulling direction}

\begin{figure}[!htbp]
\centering
\subfigure[]{\includegraphics[width=0.4\linewidth]{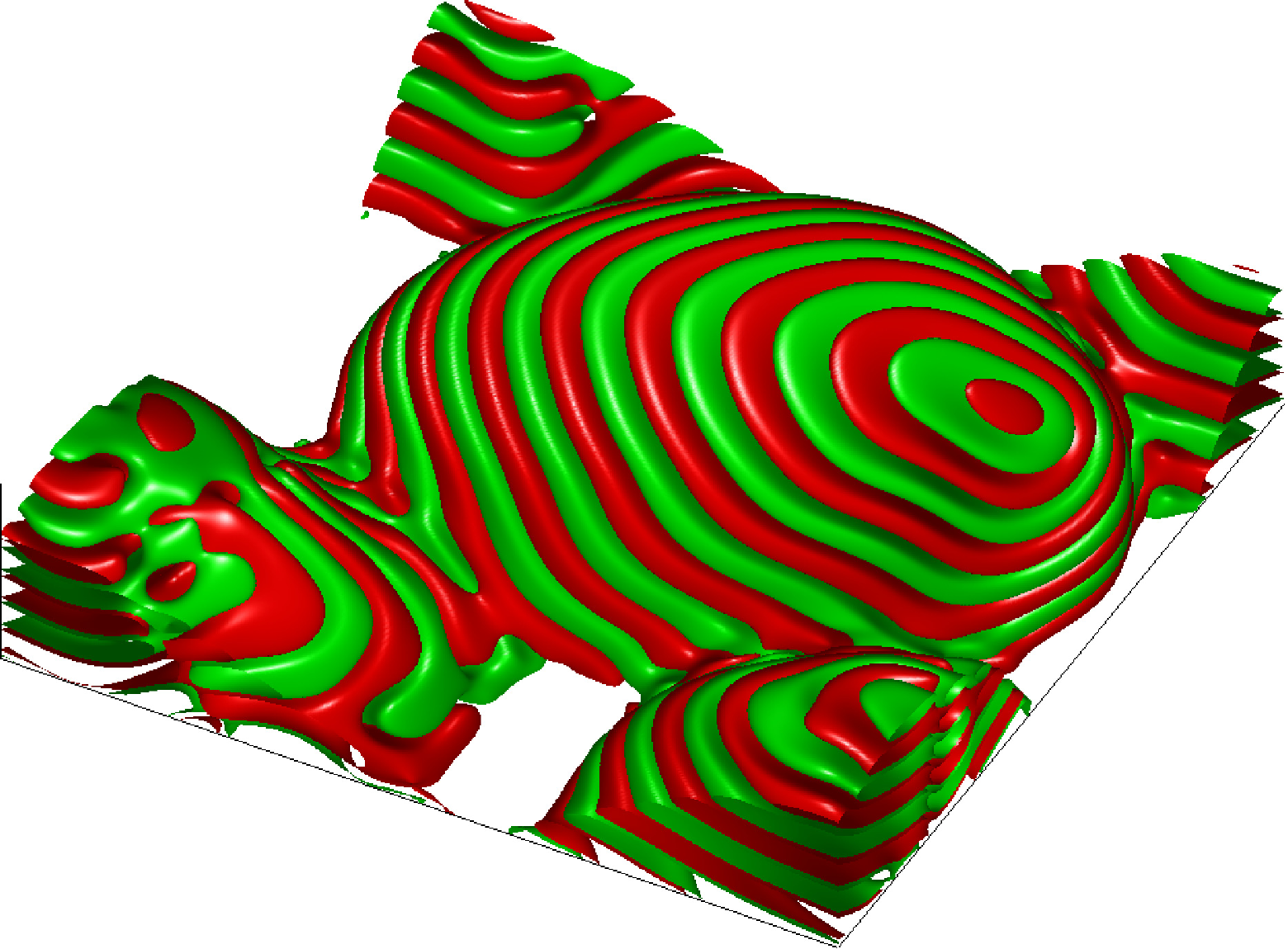}
\label{fi_y_3D_1}
}
%\subfigure[]{\includegraphics[width=0.4\linewidth]{fi_y_50000}
\subfigure[]{
\begin{tikzpicture}
  \node[anchor=south west,inner sep=0] (image) at (0,0){\includegraphics[width=0.4\linewidth]{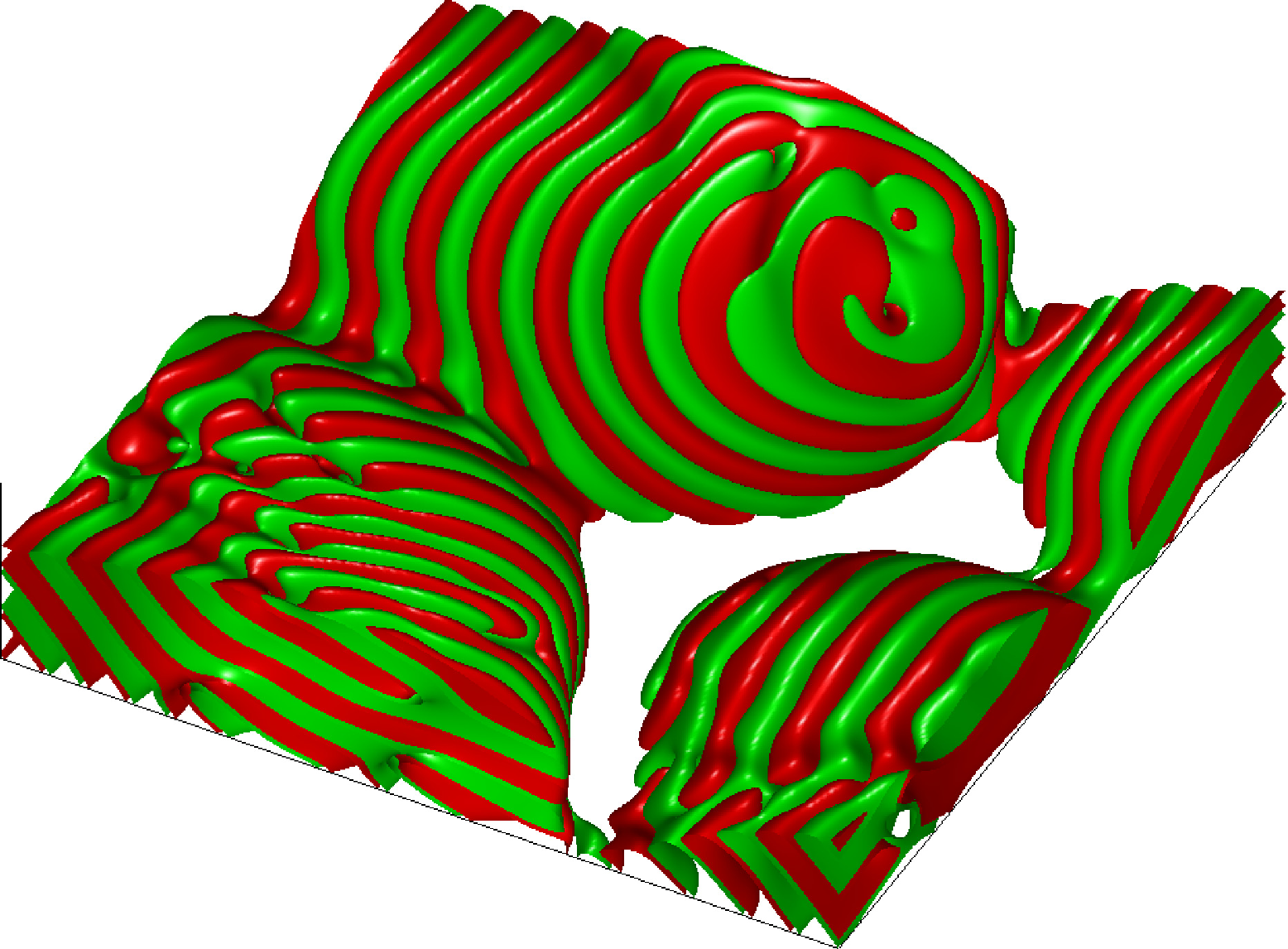}};
  \begin{scope}[x={(image.south east)},y={(image.north west)}]
   \draw [dashed,ultra thick,yellow] (0.25,0.85) --(0.95,0.5);
   \end{scope}
   \end{tikzpicture}
\label{fi_y_3D_2}
}
%\centering
\caption{Eutectic colonies at a total time of (a)22000, and (b)50000 in a system with solid-liquid interfacial energy anisotropy 
with $V=0.1$, $\delta=0.015$ and $\theta_R=10\degree$, clockwise, about an axis normal to the pulling direction. The yellow dashed line in 
(b) represents the orientation of the vertical sections reported in Fig.~\ref{fi_y_sect_y}.}
\label{fi_y_3D}
\end{figure}

Moving onto the situation where the reference frame of the crystal is rotated about an axis perpendicular 
to the direction of the imposed temperature gradient (see Fig.~\ref{fi_y_3D}), 
we find fingers taking up well-defined orientations with respect to the pulling direction. 
The eutectic spirals in this case can be seen to be traversing the simulation box in a 
direction perpendicular to the pulling direction. This is 
due to the non-zero angle to the pulling direction taken up by the fingers while growth. 
Thus, the growth velocity has a lateral component which 
creates a traveling wave of eutectic fingers across the simulation box during growth. 

\begin{figure}[!htbp]
\centering
\subfigure[]{\includegraphics[width=0.4\linewidth]{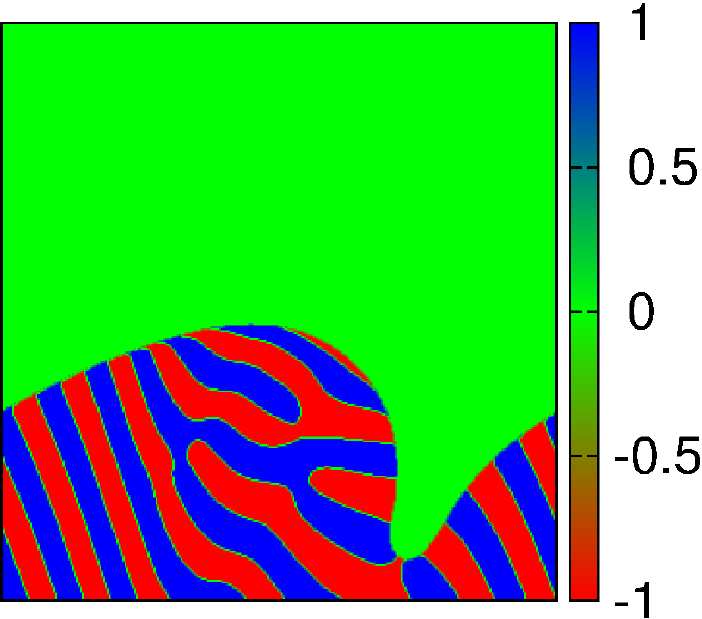}
\label{fi_y_sect_y_1}
}
%\hspace{.25in}
\subfigure[]{\includegraphics[width=0.4\linewidth]{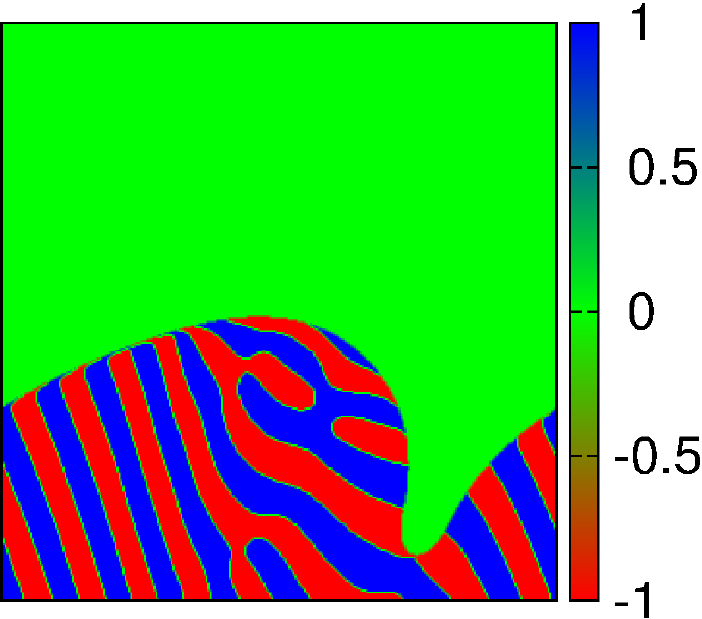}
\label{fi_y_sect_y_2}
}
\centering
\caption{2D sections ($u$ field) by a plane parallel to the pulling direction and normal to the axis of rotation of Fig.~\ref{fi_y_3D_2},
whose orientation is denoted by the yellow dotted line in Fig.~\ref{fi_y_3D_2}.
Both figures (a) and (b) highlight that the axis of the finger is not straight. Colorbars report values from the $u$ field.}
\label{fi_y_sect_y}
\end{figure}

\begin{figure}[!htbp]
\centering
\subfigure[]{\includegraphics[width=0.4\linewidth]{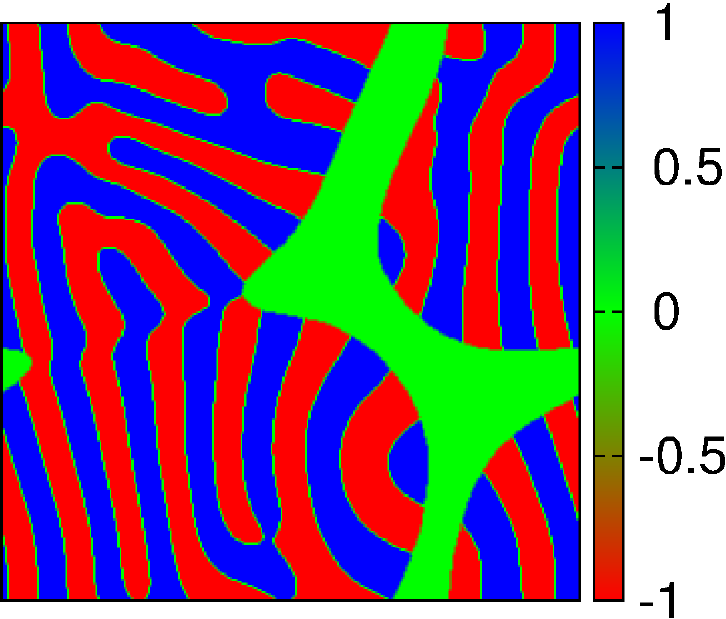}
\label{fi_y_sect_z_1}
}
%\hspace{.25in}
\subfigure[]{\includegraphics[width=0.4\linewidth]{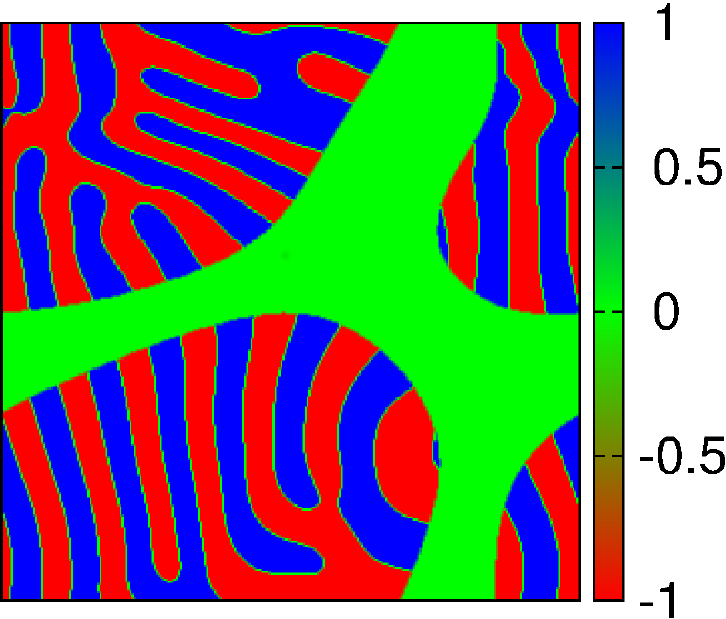}
\label{fi_y_sect_z_2}
}
%\centering
\caption{2D sections ($u$ field) by a plane normal to the pulling direction and parallel to the axis of rotation of Fig.~\ref{fi_y_3D_2}. 
The sectioning height is lower in (a) than in (b). Colorbars report values from the $u$ field.}
\label{fi_y_sect_z}
\end{figure}

Again, we consider sections of Fig.~\ref{fi_y_3D} which are parallel (Fig.~\ref{fi_y_sect_y}) and normal 
(Fig.~\ref{fi_y_sect_z}) to the 
pulling direction. The lack of a section which clearly demonstrates the axis as we have seen in 
Fig.~\ref{iso_fi_sect_y_1}, suggests that the finger axis is not completely contained in a single plane of such an orientation. 
The orientation of spirals is determined by a force balance along the tri-junction lines during its
formation via the amplification of an instability, which being an unsteady phenomenon, can lead to orientations
which deviate from the equilibrium orientation of interfacial planes.
% This is possible because we haven't specified the rotation angle to be 
% zero about the other direction which is normal to both the pulling
% direction and the current axis of rotation, which leaves the system to enjoy a degree of freedom
%while colony formation leading to introduction of indeterminacies in finger orientations. 
We can also add that, the observation of a finger axis
in Fig.~\ref{iso_fi_sect_y_1} is accidental and it could very well have been like the situation depicted here.

The sections in Fig.~\ref{fi_y_sect_z} are the ones which are taken to be perpendicular to the pulling direction and parallel to the axis of rotation.
Here the individual phases are either elongated or curved with the concavity towards the solid-liquid interface which is quite different to what
we observe for the isotropic case in Fig.~\ref{iso_fi_sect_z} and can be understood as a consequence of the tilt of the spirals. 

As we saw in 2D, the simulations done with an anisotropic solid-liquid interface leads to a stable finger width and orientation 
(see Figs.~\ref{fi_z_3D_1},~\ref{fi_z_3D_2},~\ref{fi_y_3D_1} and~\ref{fi_y_3D_2}) being selected. Thus, as opposed
to the isotropic case, the spirals once formed never disintegrate, but only split 
when they coarsen beyond the system selected finger width. 

Having understood the effect of anisotropic solid-liquid interface on the colony features in 3D, we do the 
same for solid-solid interfacial anisotropy in the following section. 

\subsection{Anisotropic solid-solid interface}
The $\alpha$-$\beta$ interface can also have preferred orientations 
with respect to the direction of the imposed 
temperature gradient, resulting in novel patterns in the eutectic colonies. The 
introduction of anisotropy is done in the same way as in 2D (see Eq.~\ref{aniso_u_fsol}); the anisotropy
function $a_c$ being given by Eq.~\ref{ac_fi_3D} with $u$'s taking the place of $\phi$'s.

\begin{figure}[!htbp]
\centering
\subfigure[]{\includegraphics[width=0.4\linewidth]{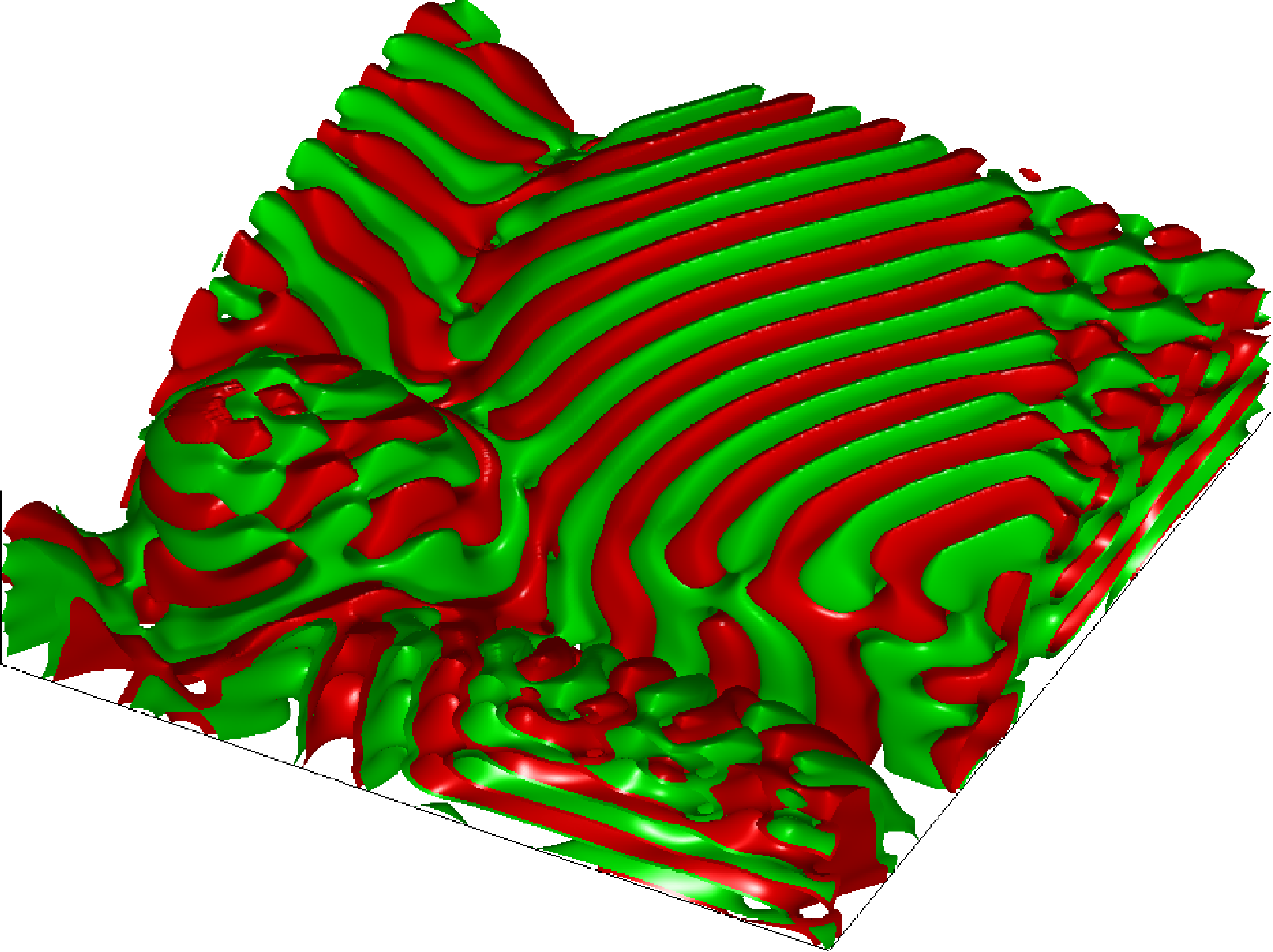}
\label{u_z_3D_1}
}
\subfigure[]{\includegraphics[width=0.4\linewidth]{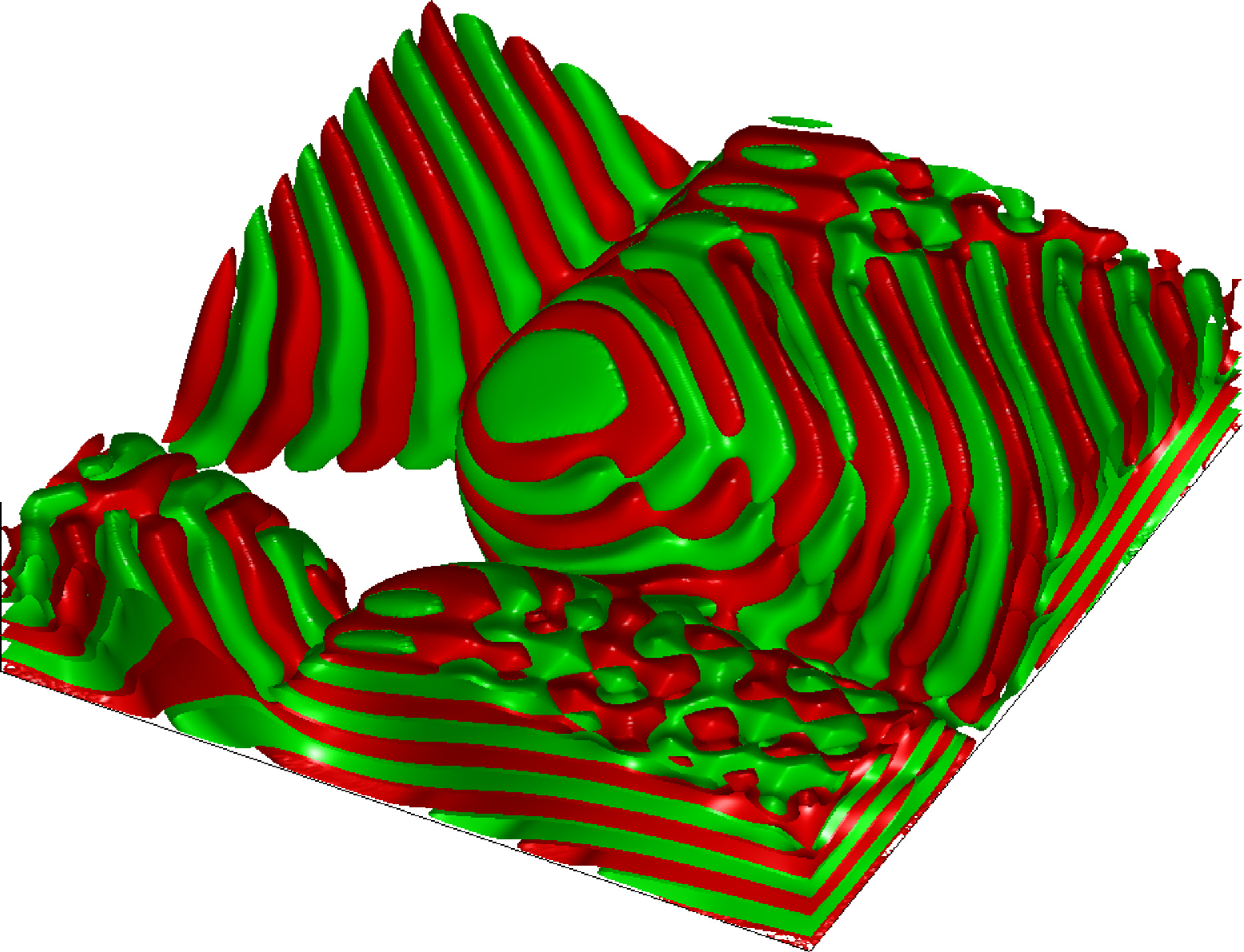}
\label{u_y_3D_2}
}
%\centering
\caption{Eutectic colonies in a system with $\alpha$-$\beta$ interfacial energy anisotropy 
with $V=0.1$, $\delta=0.015$ and $\theta_R=10\degree$, clockwise, about, (a)the pulling direction, and (b)normal to the pulling direction.}
\label{u_3D}
\end{figure}

In contrast to the stable spirals obtained for solid-liquid anisotropic interfacial energies, we do not 
get any spiraling for the crystal frame rotated about the pulling direction (Fig.~\ref{u_z_3D_1}) and only intermittent spiraling 
for the situation where the crystal frame is rotated about a normal to the pulling 
direction (Fig.~\ref{u_y_3D_2}) with the eutectic solids taking up certain well-defined orientations. 

As can be seen from Fig.~\ref{u_3D}, that the consideration of a four-fold anisotropy in the solid-solid interfacial energy leads
to individual eutectic solids to arrange themselves as alternate plates which take up orientations dictated by the anisotropy. 
The lack of a stabilizing influence of the imposed anisotropy required for forming spirals indicates 
that the equilibrium interfacial orientations do not allow the formation stable spirals. 

\section{Summary of the results}
We studied eutectic colony formation in both 2D and 
in 3D, in systems with preferred orientation of the
solid-liquid as well as the solid-solid ($\alpha$-$\beta$) interfaces. 
The 2D simulations with anisotropic solid-liquid interfacial energy
display a stable finger spacing which has a definite orientation to the pulling direction 
decided by the imposed temperature gradient
from the possibilities offered by equilibrium orientations of the solid-liquid interface 
under anisotropy. The stability of finger spacing is a function of the magnitude of anisotropy ($\delta$), with the system 
selecting a well defined finger width and tip radius only at higher $\delta$.
Higher values of $\delta$ also lead to a selection of
a more pronounced orientation of the fingers with respect 
to the pulling direction.
These observations stand in stark contrast to the isotropic 
simulation in 2D where neither a stable finger spacing 
nor a well-defined orientation is taken up by the fingers.  
At the solidification envelope, the lamellae are oriented 
normal to the solid-liquid interface (also observed in the isotropic case),
with no specific orientation at the center of the fingers. The effect of 
a higher pulling velocity ($V$) on the simulations is to modify the length scales
 manifesting as smaller lamellar width and tip radius.

When studying 2D systems with anisotropic 
solid-solid ($\alpha$-$\beta$) interfaces, we find 
a significant portion of the lamellae to assume vertical or 
near-vertical orientations in conjunction with a slight tilt of the
solid-liquid interface from the horizontal. As a consequence of this, the fingers
appear much broader than what is seen for the isotropic and solid-liquid anisotropy case.
With increase in $\delta$, the lamellae appear to favor the vertical orientation 
strongly, as more and more of them take up such orientations. Higher pulling 
velocities $V$, resulted in finer length scales in accordance to our observation
in the solid-liquid anisotropy case. 

% \textcolor{red}{The lamellar orientations from our simulations in this case also points
% to the possibility of the lamellar features in Fig.~\ref{exp1} being a 
% result of anisotropic solid-solid interfaces.}

Though eutectic spirals are observed in 3D isotropic simulations, they are not stable. 
The existing spirals disintegrate while new ones come into existence with this process repeating 
throughout the course of the simulation. 
% It should be mentioned at this point that prior to 
% our observation, spiralling in isotropic systems is observed 
% only temporarily at the tips of the fingers~\cite{Ratkai2015}.   

The 3D simulations performed with anisotropic solid-liquid interfaces, with the 
crystal frame rotated about the pulling direction, revealed spirals similar 
to the isotropic case with the solidification envelope appearing angular.
While another study done with the crystal frame rotated about an axis normal 
to the pulling direction led to tilted spirals. A stable finger width
is selected in both the simulations. 

The introduction of anisotropy in the solid-solid ($\alpha$-$\beta$) 
interfaces lead to unstable fingers in the simulations. 
The rotation of the crystal frame about the pulling direction does not create spirals, 
rather we find elongated lamellae along particular directions normal to the vertical, 
which modifies to form unstable spirals when the crystal frame is rotated about an 
axis normal to the pulling direction. 

Our 2D simulations with solid-solid anisotropy have offered a possible explanation for the existence of lamellae
 oriented along the finger axis in Fig.~\ref{exp1}. Through our 3D simulations, we are able to delineate the 
features of eutectic spirals~\cite{Akamatsu2010} for anisotropic solid-liquid and solid-solid interfaces. 
Anisotropic solid-liquid interfaces lead to the formation of stable spirals while
the choice of solid-solid anisotropy leads to unstable structures occasionally resembling spirals. 
% Simulating the secondary arm spirals as shown in Fig.~\ref{exp1}, mandates larger simulation domains and longer 
% durations of study.

\section{Conclusions}
We have attempted to understand pattern formation in eutectics in the presence of a ternary impurity under conditions of solid-liquid 
and solid-solid anisotropy. Here, through 2D simulations, we have tried to understand the effect of pulling velocity ($V$) 
and magnitude of the anisotropy ($\delta$)
on the eutectic colony microstructures for a single given rotation of the crystal frame with respect to the laboratory frame ($\theta_R$). 
The patterns possible for other $\theta_R$'s under different values of $V$ remain unexplored. The 3D simulations are carried out at a 
single value of both $\theta_R$ and $V$. The existence of novel microstructures at other combinations of these two parameters cannot be ruled 
out and stands as a promising area for further study. Furthermore, the absence of a stable spiral in 3D when considering solid-solid anisotropy,
prompts a closer look at spiral formation and conditions which allow or impede spiraling. 
Experimental studies of colony formation in anisotropic systems are critical to this end which will serve as 
a guide for choosing appropriate anisotropy functions in models.
% Also, in order to capture the spiralling secondary arms 
% seen in Fig.~\ref{exp1}, larger simulation domains and longer time iterations are required, presenting computational challenges 
% for their further exploration. 

Though we have been able to present and understand a lot of features of eutectic colonies for anisotropic 
solid-liquid and solid-solid interfaces in this study, the high solute trapping encountered for higher values of $\delta$
and $V$ in this model point towards the possibility of employing classic solidification models~\cite{Folch2005,Choudhury+11-3} 
for studying this problem to obviate this difficulty. This remains part of our future 
plans.

\section{Acknowledgements}
We are grateful to Prof. Mathis Plapp, Ecole Polytechnique, for all the insightful and critical discussions 
and thank him for his support.

\section{Appendix}
\subsection{Equilibrium compositions of the phases}
Solving the following set of non-linear equations yields the equilibrium compositions $u_s$, $u_l$, $\tilde{c_s}$ and $\tilde{c_l}$, 
\begin{align}\label{eq_val}
 &\frac{\partial f_{sol}}{\partial u}\Big|_{u_s} = \frac{\partial f_{liq}}{\partial u}\Big|_{u_l},  \nonumber  \\
 &\frac{\partial f_{sol}}{\partial \tilde{c}}\Big|_{\tilde{c_s}} = \frac{\partial f_{liq}}{\partial \tilde{c}}\Big|_{\tilde{c_l}},\nonumber \\ 
 &f_{sol}-\frac{\partial f_{sol}}{\partial u}\Big|_{u_s} u_s - \frac{\partial f_{sol}}{\partial \tilde{c}}\Big|_{\tilde{c_s}} \tilde{c_s} \nonumber \\
 &= f_{liq}- \frac{\partial f_{liq}}{\partial u}\Big|_{u_l} u_l -\frac{\partial f_{liq}}{\partial \tilde{c}}\Big|_{\tilde{c_l}} \tilde{c_l}.
\end{align}
Eqs.~\ref{eq_val} represent 
the equality of chemical potentials of $u$ and $\tilde{c}$ with an 
equality of grand potentials of the two phases as the third criterion of equilibrium.

\bibliography{eu2}

\end{document}